\documentclass[11pt,a4paper]{article}
\usepackage{jheppub}
\pdfoutput=1
\usepackage[caption=false]{subfig}
\usepackage{amsmath}
\usepackage{amsfonts}
\usepackage{amssymb}
\usepackage{float}
\usepackage{color}
\usepackage[normalem]{ulem}
\usepackage{graphicx}
\usepackage{graphics}
\usepackage{hyperref}
\usepackage{adjustbox,array}
\usepackage{enumitem} 
\hypersetup{}
\usepackage{epstopdf}
\usepackage{multirow}
\usepackage{slashed}
\usepackage{booktabs}
\usepackage{cancel}
\usepackage{mathrsfs}
\usepackage{csquotes}
\usepackage{diagbox}
\usepackage{pifont}
\definecolor{rosso}{cmyk}{0,1,1,0.4}
\definecolor{rossos}{cmyk}{0,1,1,0.55}
\definecolor{rossoc}{cmyk}{0,1,1,0.2}
\definecolor{blu}{cmyk}{1,1,0,0.3}
\definecolor{blus}{cmyk}{1,1,0,0.6}
\definecolor{bluc}{cmyk}{1,1,0,0.1}
\definecolor{verde}{cmyk}{0.92,0,0.59,0.25}
\definecolor{verdec}{cmyk}{0.92,0,0.59,0.15}
\definecolor{verdes}{cmyk}{0.92,0,0.59,0.4}

\newcommand{\eq}[1]{Eq.~(\ref{#1})}

\newcommand{\gsim}{\gtrsim}
\newcommand{\lsim}{\lesssim}

\newcommand{\lf}{\left(}
\newcommand{\ri}{\right)}

\newcommand{\nn}{\nonumber}

\newcommand{\sqt}{\sqrt{2}}
\newcommand{\hf}{h_{\rm f}}
\newcommand{\rr}{{\gamma\gamma}}

\renewcommand{\lg}{\mathscr{L}} 

\newcommand{\mco}{\mathcal{O}}

\newcommand{\br}{{\mathcal{B}}}
\newcommand{\hc}{{\rm H.c.}}

\newcommand{\znotau}{Z_{\scriptsize\mathop{\mbox{non-$\tau$}}}}

\newcommand{\sm}{{\rm SM}}

\newcommand{\fb}{{\,{\rm fb}}}
\newcommand{\ab}{{\,{\rm ab}}}
\newcommand{\ifb}{{\,{\rm fb}^{-1}}}
\newcommand{\iab}{{\,{\rm ab}^{-1}}}

\newcommand{\gev}{{\,{\rm GeV}}}
\newcommand{\tev}{{\,{\rm TeV}}}

\newcommand{\beq}{\begin{equation}}
\newcommand{\eeq}{\end{equation}}
\newcommand{\bea}{\begin{eqnarray}}
\newcommand{\eea}{\end{eqnarray}}
\newcommand{\barr}{\begin{array}}
\newcommand{\earr}{\end{array}}
\newcommand{\bc}{\begin{center}}
\newcommand{\ec}{\end{center}}
\newcommand{\bit}{\begin{itemize}}
\newcommand{\eit}{\end{itemize}}
\newcommand{\ben}{\begin{enumerate}}
\newcommand{\een}{\end{enumerate}}

\newcommand{\al}{\alpha}
\newcommand{\bt}{\beta}

\newcommand{\Dt}{\Delta}

\newcommand{\sg}{\sigma}

\newcommand{\kp}{\kappa}

\newcommand{\lmh}{\hat{\lambda}}

\newcommand{\gm}{\gamma}
\newcommand{\Gm}{\Gamma}
\newcommand{\lm}{\lambda}

\newcommand{\tauh}{{\tau_{\rm h}}}

\newcommand{\hsm}{{h_{\rm SM}}}
\newcommand{\ch}{H^\pm}

\newcommand{\wpm}{W^\pm}

\newcommand{\mh}{m_{h}}

\newcommand{\mch}{M_{H^\pm}}

\newcommand{\mhh}{M_{H}}
\newcommand{\ma}{M_{A}}
\newcommand{\mach}{M_{A/H^\pm}}
\newcommand{\mhf}{m_{h_{\rm f}}}

\newcommand{\lmc}{\Lambda_{\rm cut}}

\newcommand{\ca}{c_\alpha}
\newcommand{\sa}{s_\alpha}

\newcommand{\tb}{t_\beta}

\newcommand{\cb}{c_\beta}
\renewcommand{\sb}{s_\beta}

\newcommand{\cba}{c_{\beta-\alpha}}
\newcommand{\sba}{s_{\beta-\alpha}}

\newcommand{\ee}      {{e^+ e^-}}

\newcommand{\nnu}      {\nu\bar{\nu}}
\newcommand{\ttau}      {{\tau^+\tau^-}} 

\newcommand{\bb}      {{b \bar{b}}}

\newcommand{\qq}      {{q \bar{q}}}

\newcommand{\elll}      {{\ell^+\ell^- }}
\newcommand{\ellss}      {{\ell^\pm \ell^\pm }}

\newcommand{\met}      {{\rlap{\,/}{E}_T}}

\newcommand{\fig}[1]{Fig.~\ref{#1}} 

\newcommand{\sblue}[1]{{\color{blue} #1}} 

\newcommand{\tabincell}[2]{\begin{tabular}{@{}#1@{}}#2\end{tabular}}

\title{\color{verdes}$\boldsymbol{\tau^\pm \nu \gamma\gamma }$
 and $\boldsymbol{\ell^\pm \ell^\pm \gamma \gamma {\rlap{\,/}{E}_T} X}$ to probe the fermiophobic Higgs boson with high cutoff scales}
\author{Jinheung Kim,}
\emailAdd{jinheung.kim1216@gmail.com}
\author{Soojin Lee,}
\emailAdd{soojinlee957@gmail.com}
\author{Prasenjit Sanyal,}
\emailAdd{prasenjit.sanyal01@gmail.com}
\author{Jeonghyeon Song,}
\emailAdd{jhsong@konkuk.ac.kr}
\author{and Daohan Wang}
\emailAdd{wdh9508@gmail.com}
\affiliation{Department of Physics, Konkuk University, Seoul 05029, Republic of Korea}

\abstract{
The light fermiophobic Higgs boson $h_{\rm f}$ in the type-I two-Higgs-doublet model can evade the current search programs at the LHC since its production through the quark-antiquark annihilation and gluon fusion is not feasible. The particle can be more elusive if the model retains stability up to the Planck scale because the efficient discovery channels are missing from the existing search chart. Through the comprehensive scanning, we show that all the viable parameter points with the Planck cutoff scale require $ m_{h_{\rm f}} \in[80,\, 120]{\;{\rm GeV}}$ and $M_{A/H^\pm} \in [90,\,150]{\;{\rm GeV}}$. Since $h_{\rm f}h_{\rm f}\to \gamma\gamma W^+ W^-$ and $H^\pm \to \tau^\pm \nu/h_{\rm f}W^\pm$ are dominant in this case, two final states are more efficient to probe $h_{\rm f}$ than the conventional search mode of $4\gamma+W^\pm/Z$. One is $\tau^\pm\nu \gamma\gamma$ from $pp \to H^\pm(\to\tau^\pm\nu) h_{\rm f}(\to \gamma\gamma)$ and the other is $\ell^\pm \ell^\pm \gamma\gamma  {\rlap{\,/}{E}_T} X$ ($\ell^\pm=e^\pm,\mu^\pm$) from $pp \to H^\pm(\to h_{\rm f}W^\pm) h_{\rm f} \to \gamma\gamma W^+ W^-W^\pm $, $pp \to  H^\pm(\to h_{\rm f} W^\pm) A(\to h_{\rm f} Z) \to \gamma\gamma W^+ W^- W^\pm Z $, and $pp \to H^+(\to h_{\rm f} W^+)H^-(\to h_{\rm f} W^-)\to \gamma\gamma W^+ W^- W^+ W^-$. The inclusive $\ell^\pm \ell^\pm \gamma\gamma {\rlap{\,/}{E}_T} X$ consists of a same-sign dilepton, two prompt photons, and missing transverse energy. We perform the signal-background analysis at the detector level. With the total integrated luminosity of $300\;{\rm fb}^{-1}$ and the 5\% background uncertainty, two proposed channels at the 14 TeV LHC yield signal significances above five in the entire viable parameter space of the fermiophobic type-I with a high cutoff scale.
}

\vspace{1cm}

\begin{document}

\maketitle
\flushbottom

\section{Introduction}

The measurement of a Higgs boson with a mass of $125\gev$ at the LHC~\cite{ATLAS:2012yve,CMS:2012zhx}
is a great triumph in particle physics.
The Higgs boson is invaluable not only because it is the last piece of the standard model (SM)
but also because it shall be the stepping stone to the ultimate theory of the Universe.
The SM cannot answer
the fundamental questions such as the naturalness problem~\cite{Dimopoulos:1995mi,Chan:1997bi,Craig:2015pha}, 
the identity of dark matter~\cite{Navarro:1995iw,Bertone:2004pz},
neutrino masses, baryogenesis, the metastability of the SM vacuum~\cite{Degrassi:2012ry},
and the fermion mass hierarchy.

Nevertheless, the new physics signal that we thought would appear soon has yet to arrive.
In dealing with this disappointing situation, 
the very first task is to investigate whether we might miss 
the signal of new physics beyond the SM (BSM).
A dramatic possibility is that 
a new particle with an intermediate-mass around $100\gev$ escapes the current experimental programs.
How could it happen?
The elusiveness can come in two ways, through the suppressed production at the LHC
and the omission of efficient discovery channels from the search chart.
A light fermiophobic Higgs boson $\hf$ in the type-I two-Higgs-doublet model (2HDM)~\cite{Akeroyd:1995hg,Akeroyd:1998ui,Akeroyd:1998dt,Barroso:1999bf,Brucher:1999tx,Akeroyd:2003bt,Akeroyd:2003xi,Akeroyd:2007yh,Arhrib:2008pw,Gabrielli:2012yz,Berger:2012sy,Gabrielli:2012hd,Cardenas:2012bg,Ilisie:2014hea,Delgado:2016arn,Mondal:2021bxa,Bahl:2021str} 
is a good candidate which fits both.
The light $\hf$ with a mass below $125\gev$ is accommodated in the inverted Higgs scenario
where the heavier $CP$-even Higgs boson $H$ is the observed one~\cite{Bernon:2015wef,Chang:2015goa,Jueid:2021avn,Lee:2022gyf}.
The fermiophobic nature of $\hf$ is guaranteed by the condition of $\al=\pi/2$, 
where $\al$ is the mixing angle between two $CP$-even Higgs bosons in the 2HDM,
since all the Yukawa couplings of $\hf$ are proportional to $\cos\al$ in type-I.
The vanishing couplings to the SM fermions do not allow the production of $\hf$
via the quark-antiquark annihilation and the gluon fusion (through quark loops).
In addition, the observation of the SM-like Higgs boson~\cite{Aad:2019mbh,CMS:2020xwi,ATLAS:2021vrm}, which demands almost the Higgs alignment limit, suppresses the vector boson fusion production and the associated production of $\hf$ with a gauge boson $V(=\wpm,Z)$.
We need to resort to the associated production of $\hf$ with another BSM Higgs boson 
via electroweak processes, $pp/p\bar{p} \to \ch\hf/ A \hf$,
which have generically small cross sections.

\begin{table}
\begin{tabular}{|c||c|c|c|c|c|}
\hline
\multirow{2}{*}{\diagbox{\texttt{product}}{\texttt{final}}}& \multirow{2}{*}{$\rr\bb$} & $\rr\elll$ 
	& $4\gm \nnu$  & \multirow{2}{*}{$\rr  X$} & \multirow{2}{*}{$4\gm X$}\\
	& & $\rr\nnu,\rr qq$ &  $4\gm qq$ & & \\
\hline\hline
\multirow{2}{*}{$\ee\to A \hf$} & ~DELPHI~ & \multirow{2}{*}{}  &\multirow{2}{*}{}  & \multirow{2}{*}{}  & \multirow{2}{*}{} \\ 
& \cite{DELPHI:2001pxd,DELPHI:2003hpv} &  & & &  \\ \hline
$\ee\to A \hf$ & \multirow{2}{*}{} & \multirow{2}{*}{}   & ~DELPHI~ & \multirow{2}{*}{}  & \multirow{2}{*}{}  \\ 
$ ~~~~~~~~~\to \hf\hf Z$ & & &   \cite{DELPHI:2001pxd,DELPHI:2003hpv} &  &  \\ \hline
\multirow{2}{*}{$\ee\to \hf Z$} & \multirow{2}{*}{} & ~DELPHI~  &  \multirow{2}{*}{}  &  \multirow{2}{*}{} & \multirow{2}{*}{}  \\ 
 &  & \cite{DELPHI:2001pxd,DELPHI:2003hpv}  & & & \\ \hline \hline
$p\bar{p} \to \ch\hf $ & \multirow{2}{*}{} & \multirow{2}{*}{} &  \multirow{2}{*}{} & \multirow{2}{*}{} &\multirow{2}{*}{~CDF\cite{CDF:2016ybe}~} \\ 
$~~~~~~~\to \hf\hf W^{(*)}$ &&&&&\\ \hline
$pp\to \hsm \to \hf\hf$ & & & & &  CMS\cite{CMS:2021bvh}\\[1pt] \hline
\multirow{4}{*}{$pp/p\bar{p} \to \hf V/\hf jj$} &  \multirow{4}{*}{} & \multirow{4}{*}{} & \multirow{4}{*}{} & CDF~\cite{CDF:2009pne}& \multirow{4}{*}{} \\ 
 &  & & &  D0~\cite{D0:2008srr,D0:2011kow}& \\ 
  &  & & &  CMS~\cite{CMS:2012mua,CMS:2012xok,CMS:2013zma}& \\
  &  & & &  ATLAS~\cite{ATLAS:2012yxc}& \\ \hline
\end{tabular}
\caption{\label{table:exp}
Experimental searches for a fermiophobic Higgs boson
at the LEP, Tevatron, and LHC.
We classify the processes according to the production channel and the final states.}
\end{table}

The second loophole, the omission of the efficient discovery channels from the search chart,
also happens to the fermiophobic Higgs boson $\hf$ in type-I.
The searches for $\hf$ up to now are summarized in Table \ref{table:exp},
classified according to the production processes and the final states.
The experimental searches are based on the assumption of
$\br(\hf\to\rr)\sim 100\%$ and $\br(\ch\to \hf \wpm)\sim 100\%$.
On the theoretical side,
the final states of $4\gm+V$~\cite{Akeroyd:2003bt,Akeroyd:2003xi,Akeroyd:2005pr,Delgado:2016arn,Arhrib:2017wmo}
and $4\gm+VV'$~\cite{Kim:2022nmm} have been mainly studied under the same assumption.
In a large portion of the viable parameter space, however, $\hf\to\rr$ is not the leading one
because of the sizable three-body and four-body decays of $\hf \to W^{(*)}W^*$.
If $\mch\lsim \mhf+15\gev$, furthermore, the charged Higgs boson decays dominantly into $ \tau^\pm\nu$.
In this case, the process of $pp/p\bar{p} \to \hf\ch(\to \hf \wpm) \to 4\gm \wpm$
is not efficient.
Although $\br(\hf\to\rr)\simeq 100\%$ is achievable when $\mhf$ is very light (below about $30\gev$),
the fermiophobic type-I with such a light $\hf$ is not well-motivated in two respects.
First, the phenomenologically allowed parameter space is extremely limited:
only about 0.01\% of the parameters\footnote{The scanning ranges are $\mach \in [80,900]\gev$, $\tan\beta\in [1,100]$, and $m_{12}^2 \in [0,15000]\gev^2$. Here $\tan\beta$ is the ratio of two vacuum expectation values of two Higgs doublet fields
and $m_{12}^2$ is a soft breaking parameter of $Z_2$ parity,
both of which are to be defined below.}
 that meet the theoretical requirements satisfy the experimental constraints~\cite{Kim:2022nmm}. 
Second, the light $\hf$ entails a low cutoff scale.
As the parameters evolve under the renormalization group equations (RGEs),
theoretical stability (perturbativity, unitarity, or vacuum stability)
is broken at the cutoff scale $\lmc$~\cite{Machacek:1983tz,Machacek:1983fi,Machacek:1984zw,Haber:1993an,Luo:2002ti,Grimus:2004yh,Chakrabarty:2014aya,Das:2015mwa,Chakrabarty:2016smc,Kang:2022mdy}.
If $\mhf$ is light enough to guarantee $\br(\hf\to\rr)\simeq 100\%$,
$\lmc$ is low. 
For example, the maximum $\lmc$ for $\mhf=30\gev$ is only $7\tev$~\cite{Kim:2022nmm},
which calls for an extension of the model 
to include other heavy BSM particles with masses about $\mco(1)\tev$.
If $\mhf\gsim 80\gev$, however,
the model can retain stability up to the Planck scale.
While we are not claiming that the fermiophobic type-I is the ultimate theory, 
we do not anticipate either that the model will be superseded at a significantly low energy scale.
Most of all, the viable parameter points with $\mhf\gsim 80\gev$ 
do not incorporate $\br(\hf\to\rr)\sim100\%$,
implying efficient discovery channels other than $4\gm +V$.
The parameter space with a high cutoff scale should be meticulously explored before devising an extension of the model.

Starting from the given motivation, 
we aim to investigate the phenomenological properties of the fermiophobic type-I 2HDM 
with  $\lmc>10^{18}\gev$. 
The parameter space with such a high cutoff scale exhibits certain characteristics 
that suggest $\tau^\pm\nu\rr$ and $\ellss\rr\met X$ ($\ell^\pm = e^\pm,\mu^\pm$) 
as the most efficient channels to probe $\hf$ at the LHC.
Here, the notation $\ellss\rr\met X$ refers to the inclusive final state 
consisting of a pair of same-sign leptons, 
a pair of photons, and missing transverse energy. 
The two channels have not been studied in the literature and thus merit a thorough exploration.
Later on, we will relax the constraint of $\lmc>10^{18}\gev$ and show that these two channels continue to be efficient in probing $\hf$ for a substantial portion of the parameter space even with a lower cutoff scale.
An essential prerequisite
is the preparation of the parameter points that satisfy the theoretical and experimental constraints.
Although most of the direct search bounds are covered by the open code \textsc{HiggsBounds}~\cite{Bechtle:2020pkv},
the search for $\hf$ in the mode of $4\gm X $ by the CDF Collaboration~\cite{CDF:2016ybe}
is missing.
In addition, the measurement of $W(\to\ell\nu)\rr$ by the CMS Collaboration~\cite{CMS:2021jji}
constrains the model because the final state of $\ell\nu\rr$
occurs from the signal of $pp\to\hf(\to\rr)\ch(\to \tau^\pm\nu)$ 
followed by the leptonic decay of $\tau^\pm$.
We will impose the two constraints.
Based on the viable parameter points at the electroweak scale,
we will run the parameters under the RGEs and acquire the parameter points with a high cutoff scale.
It is to be shown that the condition of $\lmc>10^{18}\gev$ has big impacts 
such that the dominant decay modes are $\hf\to WW^*/\rr$, 
$\ch\to \tau^\pm\nu/\hf W^*$, and $A\to \hf Z$.
Then $\tau^\pm\nu\rr$ and $\ellss\rr\met X$ become more efficient than $4\gm+V$.
We will additionally perform signal-background analyses at the detector level.
With the total integrated luminosity of $300\ifb$ and the 5\% background uncertainty, 
the two channels yield a signal significance above five in the entire parameter space of the model.
These are our new contributions.

The paper is organized in the following way. 
In Sec.~\ref{sec:review}, we briefly review the light fermiophobic Higgs boson in the type-I 2HDM.
Section \ref{sec:viable} describes the scanning method to impose the theoretical requirements
and the experimental constraints. 
After calculating the cutoff scale, 
we study the characteristics of the parameter points with high $\lmc$.
The branching ratios of the BSM Higgs bosons are to be also studied.
In Sec.~\ref{sec:channels},
we will discuss all the possible final states to probe $\hf$ at the LHC
and suggest two discovery channels of $\tau^\pm\nu\rr$ and $\ellss \rr \met X$.
The parton-level cross sections over the viable parameter points are presented. 
Section \ref{sec:signal:background} is devoted to the signal-background analysis.
Finally, we conclude in Sec.~\ref{sec:conclusions}.

\section{Light fermiophobic Higgs boson in type-I}
\label{sec:review}

The 2HDM introduces two \textit{SU}$(2)_L$ complex scalar doublet fields with hypercharge $Y=+1$~\cite{Branco:2011iw}:
\bea
\label{eq:phi:fields}
\Phi_i = \left( \begin{array}{c} w_i^+ \\[3pt]
\dfrac{v_i +  \rho_i + i \eta_i }{ \sqrt{2}}
\end{array} \right), \quad (i=1,2)
\eea
where $v_{1}$ and $v_2$ are the vacuum expectation values of $\Phi_{1}$ and $\Phi_2$, respectively.
The combination $v =\sqrt{v_1^2+v_2^2}=246\gev $ spontaneously breaks
the electroweak symmetry.
We define $\tan\beta= v_2/v_1$.
For simplicity, 
we use the notations of $s_x =\sin x$, $c_x = \cos x$, and $t_x = \tan x$ in what follows.

To prevent the  flavor-changing neutral currents at the tree level,
we introduce a discrete $Z_2$ symmetry,
under which $\Phi_1 \to \Phi_1$ and $\Phi_2 \to -\Phi_2$~\cite{Glashow:1976nt,Paschos:1976ay}.
With $CP$-invariance and softly broken $Z_2$ symmetry, the scalar potential is
\begin{align}
\label{eq:VH}
V_\Phi &=  m^2 _{11} \Phi^\dagger_1 \Phi_1 + m^2 _{22} \Phi^\dagger _2 \Phi_2
-m^2 _{12} ( \Phi^\dagger_1 \Phi_2 + \hc) \\ \nn
& + \frac{1}{2}\lambda_1 (\Phi^\dagger _1 \Phi_1)^2
+ \frac{1}{2}\lambda_2 (\Phi^\dagger _2 \Phi_2 )^2
+ \lambda_3 (\Phi^\dagger _1 \Phi_1) (\Phi^\dagger _2 \Phi_2)\\ \nn
& + \lambda_4 (\Phi^\dagger_1 \Phi_2 ) (\Phi^\dagger _2 \Phi_1) 
+ \frac{1}{2} \lambda_5
\left[
(\Phi^\dagger _1 \Phi_2 )^2 +  \hc
\right].
\end{align}
The model has five physical Higgs bosons, the lighter $CP$-even scalar $h$,
the heavier $CP$-even scalar $H$, the $CP$-odd pseudoscalar $A$,
and a pair of charged Higgs bosons $H^\pm$.
The relations between the mass eigenstates 
and the weak eigenstates via two mixing angles of $\al$ and $\bt$ are referred to Ref.~\cite{Song:2019aav}.
The SM Higgs boson is a linear combination of $h$ and $H$, given by
\bea
\label{eq:hsm}
\hsm = \sba h + \cba H.
\eea
To accommodate a light fermiophobic Higgs boson, 
we adopt the inverted Higgs scenario where $\mhh=125\gev$.
Considering that the observed Higgs boson at the LHC~\cite{ATLAS:2020fcp,ATLAS:2020bhl,CMS:2020zge,ATLAS:2021nsx,CMS:2021gxc,ATLAS:2020syy,ATLAS:2021upe,ATLAS:2020pvn,CMS:2021ugl,ATLAS:2020wny,ATLAS:2020rej,ATLAS:2020qdt,ATLAS:2020fzp,CMS:2020xwi,ATLAS:2021zwx} has agreed with the predictions for the SM Higgs boson up to now,
we expect $\hsm \simeq H$ (i.e., $\cba\simeq 1$),
which is dubbed the Higgs alignment limit.

The Yukawa interactions of the SM fermions are parametrized by
\begin{align}
\nn
\lg^{\rm Yuk} =
& - \sum_f 
\lf 
\frac{m_f}{v} \xi^h_f \bar{f} f h + \frac{m_f}{v} \kp_f^H \bar{f} f H
-i \frac{m_f}{v} \xi_f^A \bar{f} \gm_5 f A
\ri
\\[3pt] \nn &
- 
\left\{
\dfrac{\sqrt{2}}{v } \overline{t}
\left(m_t \xi^A_t \text{P}_- +  m_b \xi^A_b \text{P}_+ \right)b  H^+
+\dfrac{\sqt m_\tau}{v}\xi^A_\tau \,\overline{\nu}_\tau P_+ \tau H^+
+\hc
\right\}.
\end{align}
In type-I, the Yukawa coupling modifiers are
\begin{align}
\label{eq:Yukawa:couplings}
\xi^h_f&=\frac{\ca}{\sb} ,
\quad
\kp^H_f = \frac{\sa}{\sb} ,
\quad
\xi^A_t = -\xi^A_b = -\xi^A_\tau = \frac{1}{\tb}.
\end{align}

The lighter $CP$-even Higgs boson $h$ becomes fermiophobic if $\al=\pi/2$.
The fermiophobic condition can be preserved at the loop level
by a suitable renormalization condition~\cite{Barroso:1999bf,Brucher:1999tx}.
The condition of $\al=\pi/2$ yields 
\bea
\label{eq:tb}
\tb = -\frac{\cba}{\sba},
\eea
which implies $\sba<0$.
Since the observation of the SM-like Higgs boson demands almost the Higgs alignment limit ($\cba \simeq 1$),
the fermiophobic limit incorporates large $\tb$.
We summarize our model as
\bea
\label{eq:model:def}
\hbox{fermiophobic type-I:} \quad \mhh=125\gev,\quad \al=\pi/2.
\eea
In what follows,
$\hf$ denotes the lighter $CP$-even Higgs boson satisfying $\xi^h_f=0$.

The BSM Higgs bosons ($\hf$, $A$, and $\ch$) contribute to the electroweak precision observables
which are effectively parameterized by the Peskin-Takeuchi oblique parameters, $S$, $T$, and $U$~\cite{Peskin:1991sw}.
It is well known that if any two BSM Higgs bosons have the same masses,
the most sensitive parameter $T$ vanishes as in the SM~\cite{Kanemura:2011sj,Funk:2011ad,Haller:2018nnx,Lee:2022gyf}.
To efficiently satisfy the current best-fit results for the Peskin-Takeuchi oblique parameters~\cite{Peskin:1991sw}, 
we impose the mass degeneracy condition of $\ma=\mch\equiv \mach$.
Then the model has four parameters:
\bea
\label{eq:model:parameters}
\{ \mhf, \; \mach,\; m_{12}^2,\; \tb \}.
\eea
The quartic couplings in \eq{eq:VH} are written as
\begin{align}
\label{eq:quartic}
\lm_1 &= \frac{1}{v^2} \left[
\mhf^2 +\tb^2 \lf \mhf^2-M^2 \ri
\right], \\[4pt] \nn
\lm_2 &= \frac{1}{v^2} \left[
\mhh^2 + \frac{1}{\tb^2} \lf \mhh^2-M^2 \ri
\right], \\[4pt]  \nn
\lm_3 &= \frac{1}{v^2} \left[
2\mach^2-M^2
\right], \\[4pt]  \nn
\lm_4 &= \lm_5 = \frac{1}{v^2} \left[
M^2 - \mach^2
\right], 
\end{align}
where $M^2 = m_{12}^2/(\sb\cb)$.

The trilinear Higgs vertex $\hf$-$H^+$-$H^-$ plays an important role in the decay of $\hf\to \rr$.
For the scalar interaction defined by
$ \lg_{\rm trilinear} \supset  \lmh_{hH^+ H^-} \,v \hf H^+ H^- $,
the dimensionless trilinear Higgs coupling is
\begin{align}
\label{eq:hH+H-}
 \lmh_{hH^+ H^-} &= \frac{1}{v^2}
 \left[
 2 \mch^2 \cb + \mhf^2 \lf \frac{1}{\cb} - \cb \ri - \frac{1}{\cb}M^2 
 \right] 
 \\[4pt] \nn
&\simeq
\frac{\tb (\mhf^2-M^2)}{v^2} \left[ 1+\mco \lf \frac{1}{\tb^2} \ri \right] .
\end{align}
In the large $\tb$ limit, $ \lmh_{hH^+ H^-}$ is enhanced unless $\mhf^2 \approx M^2$.

\section{Scanning method and the characteristics of the high cutoff scale}
\label{sec:viable}

\subsection{Basic scanning for the viable parameter points with $\lmc>10^{18}\gev$}
\label{subsec:basic}

An essential preliminary work for the RGE analysis is the preparation of the parameter points
allowed at the electroweak scale.
In Ref.~\cite{Kim:2022nmm}, we showed that the parameter space
for $\mhf=20,30,40,60\gev$ is extremely narrow.
Under the random scan over $\mach \in [80,900]\gev$, $\tan\beta\in [1,100]$, and $m_{12}^2 \in [0,15000]\gev^2$,
only $\mco(0.01)\%$ of the parameter points 
that comply with the theoretical requirements
explain the null results in the direct searches at the LEP, Tevatron, and LHC.
In addition, the allowed parameter points have a very low cutoff scale:
for $\mhf=30\gev$, $\lmc<7\tev$.
Since our main purpose is to study the fermiophobic type-I with a high cutoff scale,
we consider
\beq
\label{eq:mhf}
\mhf=80,~90,~100\gev.
\eeq
The other parameters are randomly scanned over 
\begin{align}
\label{eq:scan:range}
& 
\mach \in \left[ 15,\, 900 \right] \gev,  \\ \nn
& \tb  \in \left[ 1,\,100 \right], \qquad
m_{12}^2 \in \left[ 0, 20000 \right] \gev^2.
\end{align}
We take only the positive values of $m_{12}^2$ in the full scanning,
because the preliminary scanning shows that none of the parameter points with negative $m_{12}^2$
can satisfy the theoretical requirements, especially the vacuum stability.
Over the random parameter points in \eq{eq:scan:range}, 
we cumulatively impose the following constraints:
\begin{description}
\item[Step A.]Theoretical requirements and the low energy data
\renewcommand\labelenumi{(\theenumi)}	
	\ben
	\item The theoretical requirements consist of the Higgs potential being bounded from below~\cite{Ivanov:2006yq},
the tree-level unitarity of scalar-scalar scatterings~\cite{Branco:2011iw,Arhrib:2000is},
the perturbativity of Higgs quartic couplings~\cite{Chang:2015goa}, and the stability of the vacuum~\cite{Ivanov:2008cxa,Barroso:2012mj,Barroso:2013awa}.
The public code \textsc{2HDMC}~\cite{Eriksson:2009ws} is used to check the requirements except for the vacuum stability.
Since \textsc{2HDMC} does not examine whether the SM vacuum is the global minimum,
we further demand the tree-level vacuum stability condition of~\cite{Barroso:2013awa}
 \bea
 \label{eq:vacuum:stability}
m_{12}^2 \left[ m_{11}^2 - \lf \frac{\lm_1}{\lm_2} \ri^{1/2} m_{22}^2 \right] 
\left[ \tb-\lf \frac{\lm_1}{\lm_2} \ri^{1/4}\right] >0,
\eea
which is sufficient  near the Higgs alignment limit~\cite{Basler:2017nzu,Branchina:2018qlf}.
	\item We demand the Peskin-Takeuchi oblique parameters of $S$, $T$, and $U$ in the 2HDM~\cite{He:2001tp,Grimus:2008nb}
	to satisfy the current best-fit results\footnote{If we accept the recent CDF measurement 
of the $W$-boson mass~\cite{CDF:2022hxs}, $m_W^{\rm CDF} = 80.4335 \pm 0.0094\gev$,
the oblique parameters change into $S_{\rm CDF}=0.15\pm 0.08$ and $T_{\rm CDF}=0.27\pm 0.06$ with $U=0$~\cite{Lu:2022bgw}.
Although $m_W^{\rm CDF}$ requires sizable mass differences among the BSM Higgs bosons in the 2HDM~\cite{Fan:2022dck,Zhu:2022tpr,Lu:2022bgw,Zhu:2022scj,Song:2022xts,Bahl:2022xzi,Heo:2022dey,Babu:2022pdn,Biekotter:2022abc,Ahn:2022xeq,Han:2022juu,Arcadi:2022dmt,Ghorbani:2022vtv,Broggio:2014mna,Lee:2022gyf,Kim:2022hvh},
our conclusions do not significantly change. }
 at 95\% C.L.~\cite{ParticleDataGroup:2022pth}:
	\begin{align}
	\label{eq:STU:PDG}
	S &= -0.02 \pm 0.10,
	\\ \nn
	T &= 0.03 \pm 0.12, \quad 
	U=0.01 \pm 0.11, 
	\\ \nn
	 \rho_{ST} &= 0.92, \quad \rho_{SU}=-0.80,\quad \rho_{TU}=-0.93,	
	 \end{align}
	where $\rho_{ij}$ is the correlation matrix. 
	In practice, our assumption of the mass degeneracy between $\ch$ and $A$ guarantees to satisfy the oblique parameters.
	\item We require that the measurements of the inclusive $B$-meson decay into
$X_s \gm$
	 should be satisfied at 95\% C.L.~\cite{Arbey:2017gmh,Sanyal:2019xcp,Misiak:2017bgg}.
	\een
\item[Step B.]   High energy collider data
\renewcommand\labelenumi{(\theenumi)}
	\ben
	\item We check the consistency with the Higgs precision data.
	The open code \textsc{HiggsSignals}-v2.6.2~\cite{Bechtle:2020uwn} is used,
	which gives the $\chi^2$ value for 111 Higgs
observables~\cite{Aaboud:2018gay,Aaboud:2018jqu,Aaboud:2018pen,Aad:2020mkp,Sirunyan:2018mvw,Sirunyan:2018hbu,CMS:2019chr,CMS:2019kqw}.
	We demand that the parameter points be within $2\sg$ confidence intervals
	with respect to the best-fit point of the fermiophobic type-I:
	the $\chi^2_{\rm min}$ value is 90.59, 90.25, 89.99 in the case of $\mhf=80,90,100\gev$ respectively,
	while in the SM $\lf \chi^2_{\rm min} \ri_\sm = 91.18$.
	\item We examine if the parameter point is consistent with the null results in the direct searches at the LEP, Tevatron, and LHC.
	The open code \textsc{HiggsBounds}-v5.10.2~\cite{Bechtle:2020pkv} is used.
	We accept only the parameter point of which the predicted cross section is smaller than the 95$\%$ C.L. upper 
	bound on the observed cross section.
	\een
	\end{description}

Over the parameter points allowed by Step A and Step B,
we run the following parameters  by using the public code \textsc{2HDME}~\cite{Oredsson:2018yho,Oredsson:2018vio}:
\bea
\label{eq:running:parameters}
g_{1,2,3}, \quad \lm_{1,\cdots,5}, \quad Y^{\varphi}_f,\quad m_{11}, \quad m_{12},
\quad m_{22}^2, \quad v_{1,2}, 
\eea
where $\varphi=\hf,H,A,\ch$ and $Y^{\varphi}_f$ denotes the Yukawa coupling defined in Ref.~\cite{Branco:2011iw}.
We set the initial energy scale to be the top quark pole mass, $m_{t}^{\rm pole} = 173.4\gev$.
The boundary conditions at $m_{t}^{\rm pole}$ are referred to Ref.~\cite{Oredsson:2018yho}.
For the evolved parameters at the next high-energy scale,\footnote{To cover the energy scale from $m_{t}^{\rm pole}$ to the Planck scale, we take a uniform step in $\log(Q)$.}
we check three conditions, tree-level unitarity, perturbativity,\footnote{Note that \textsc{2HDME}
adopts the perturbativity condition of $|\lm_{1,\cdots,5}|<4\pi$.} and vacuum stability.
If all the conditions are satisfied,
we run the parameters into the next-level high-energy scale and check the three conditions again.
If any of the three conditions is violated,
we stop the running and record the energy scale as the cutoff scale $\lmc$ for the parameter point.

\begin{figure}[!t]
\centering
\includegraphics[width=1.02\textwidth]{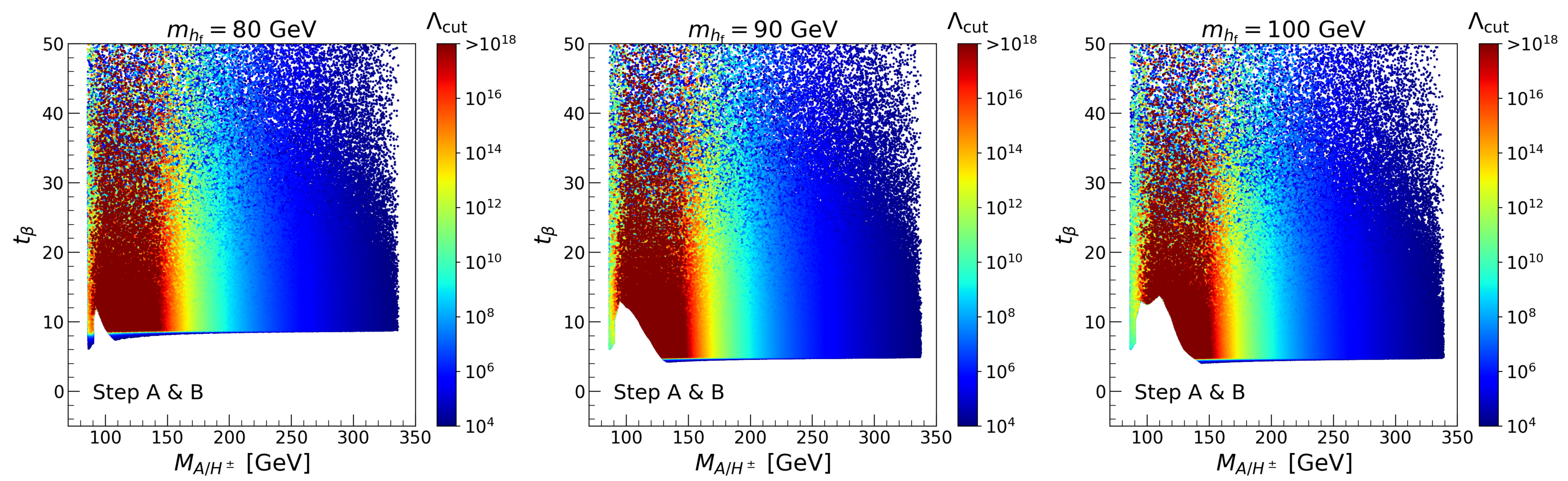}
\vspace{-0.4cm}
\caption{
$\tb$ versus $\mach$ of the parameter points that satisfy
the constraints at Step A, Step B, and $\lmc>10\tev$.
We present the results for $\mhf=80\gev$ in the left panel,
$\mhf=90\gev$ in the middle panel, and $\mhf=100\gev$ in the right panel.
The color codes denote the cutoff scale in units of GeV.
  }
\label{fig-tb-MA-cutoff}
\end{figure}

In \fig{fig-tb-MA-cutoff},
we present $\tb$ versus $\mach$ after Steps A and  B with the condition of $\lmc>10\tev$.
The results for $\mhf=80,90,100\gev$ are in the left, middle, and right panels, respectively.
The color codes denote the cutoff scale in units of GeV.
The first noteworthy feature is that the condition of $\lmc>10\tev$
already restricts the model considerably.
The BSM Higgs bosons cannot be too heavy.
The second important feature is that a large portion of the parameter space 
can retain theoretical stability up to the Planck scale.
Finally, the value of $\tb$ is also restricted.
The region with $\tb\lsim 6$ ($\tb\lsim 4$)
is excluded in the case of $\mhf=80\gev$ ($\mhf=90,100\gev$),
mainly by the LHC searches for the light charged Higgs boson~\cite{ATLAS:2018gfm,CMS:2019bfg,Sanyal:2019xcp}.

\begin{figure}[!h]
\centering
\includegraphics[width=\textwidth]{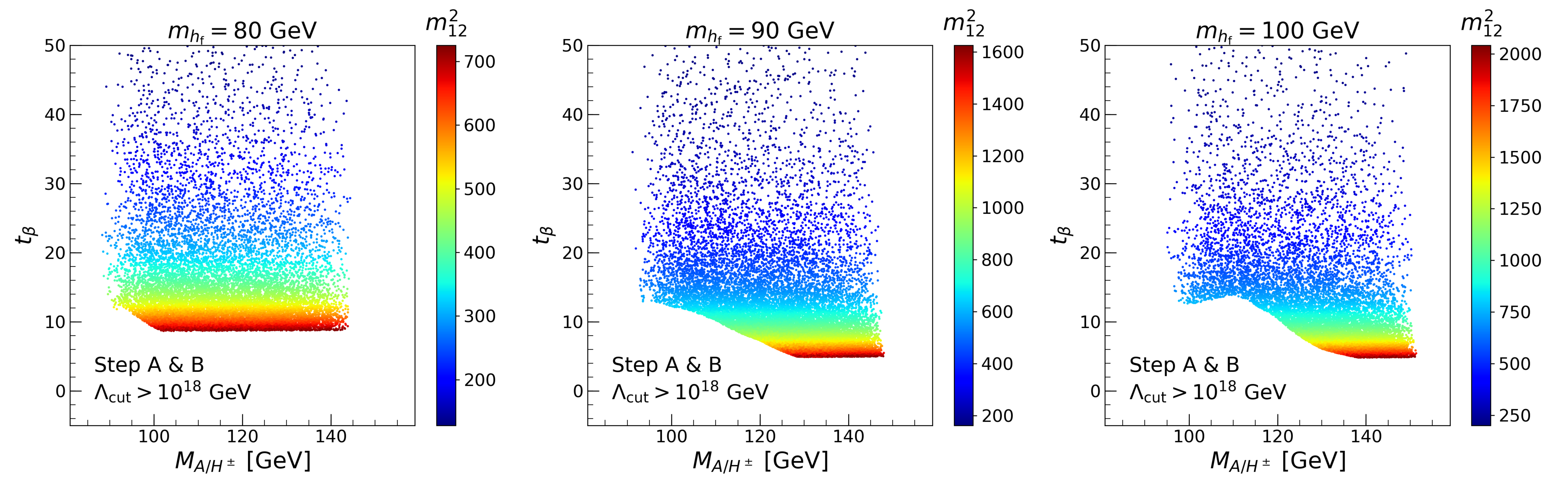}
\vspace{-0.4cm}
\caption{
$\tb$ versus $\mach$ for the parameter points with $\lmc>10^{18}\gev$.
The color codes denote $m_{12}^2$ in units of ${\rm GeV}^2$.
The results for $\mhf=80,90,100\gev$ are in the left, middle, and right panel, respectively.
  }
\label{fig-tb-MA-m12sq-Planck-AB}
\end{figure}

Let us investigate the characteristics of the allowed parameter points with the Planck cutoff scale.\footnote{The viable parameter
space with the GUT cutoff scale, $\lmc>10^{16}\gev$, is approximately the same as that with the Planck cutoff scale.}
In \fig{fig-tb-MA-m12sq-Planck-AB}, we present
$\tb$ versus $\mach$ for the parameter points that satisfy Step A, Step B, and $\lmc>10^{18}\gev$,
where the color codes denote $m_{12}^2$ in units of ${\rm GeV}^2$.
The most salient consequence from $\lmc>10^{18}\gev$
is that the other BSM Higgs bosons are also light such that $\mach\lsim 145, 148,152\gev$ for $\mhf=80, 90, 100\gev$, respectively.
The light charged Higgs boson in type-I has recently drawn a lot of interest~\cite{Arhrib:2017wmo,Akeroyd:2018axd,Cheung:2022ndq,Bhatia:2022ugu}.
As $\mach$ is similar to $\mhf$, 
the high cutoff scale requires the nearly compressed mass spectra of the BSM Higgs bosons.
On the other hand, 
the allowed values for $\tb$ remain almost the same after imposing $\lmc>10^{18}\gev$.

\begin{figure}[!h]
\centering
\includegraphics[width=\textwidth]{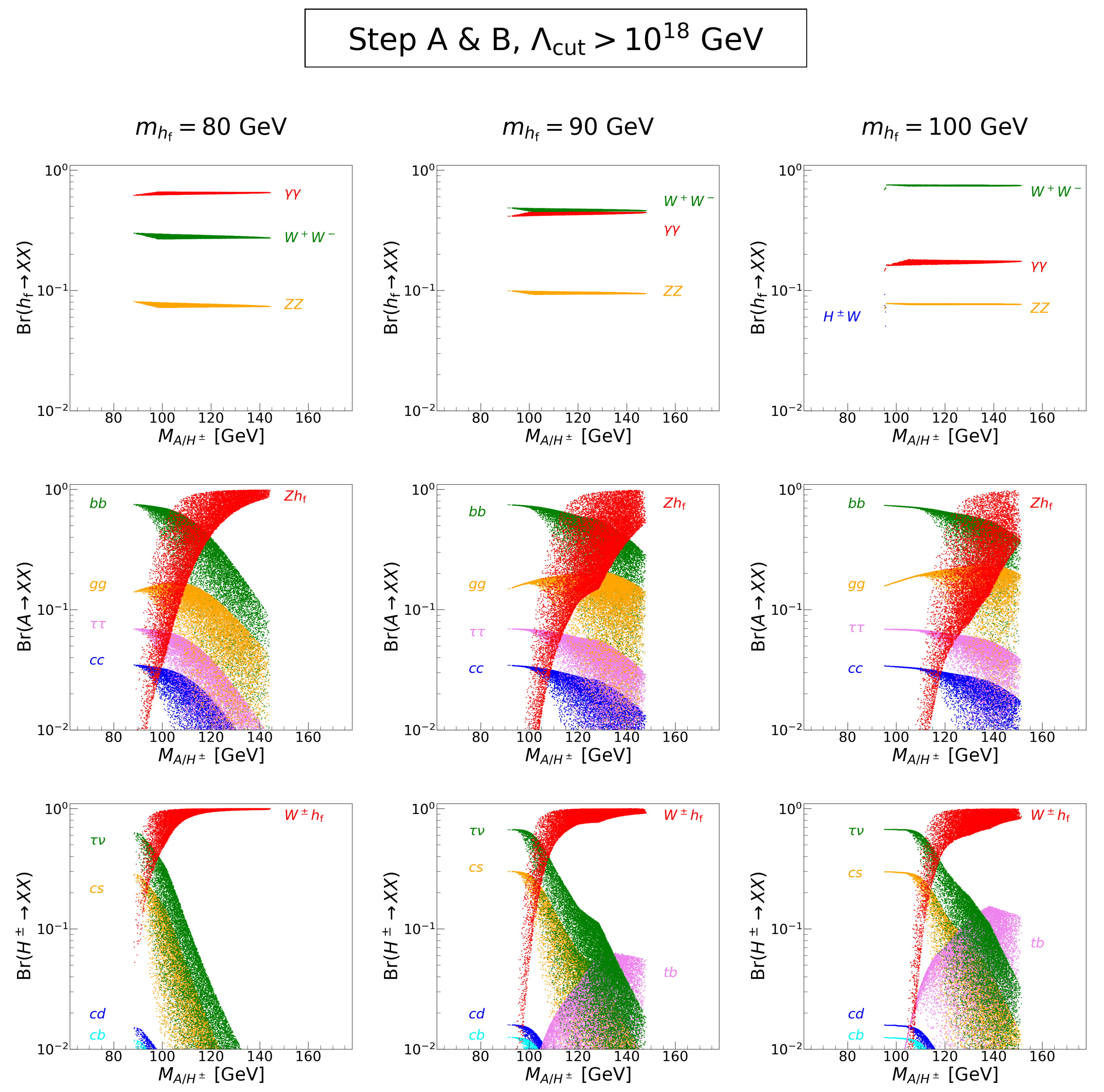} 
\caption{
Branching ratios of $\hf$ (upper panels), $A$ (middle panels), and $\ch$ (lower panels)
for the viable parameter points with $\lmc>10^{18}\gev$.
The results of $\mhf=80\gev$,  $\mhf=90\gev$, and $\mhf=100\gev$
are in the left, middle, and right panels, respectively.}
\label{fig-BR}
\end{figure}

In order to see how the BSM Higgs bosons decay if $\lmc>10^{18}\gev$,
we present in \fig{fig-BR} the branching ratios of $\hf$ (upper panels), $A$ (middle panels), and $\ch$ (lower panels)
about $\mach$, by using the \textsc{2HDMC}.
The results of $\mhf=80,\,90,\,100\gev$
are in the left, middle, and right panels, respectively.
Let us briefly review the calculation of the branching ratios in the \textsc{2HDMC}.
First, three-body decays (e.g., $\hf \to WW^* \to W f\bar{f}'$) and four-body decays (e.g., $\hf \to W^* W^* \to f\bar{f}' f\bar{f}'$)~\cite{Harlander:2013qxa} are appropriately included.
For $\ch/A/\hf\to q \bar{q}\,'$,
\textsc{2HDMC} incorporates the QCD radiative corrections at order $\alpha^2_s$
in the $\overline{\rm MS}$ scheme~\cite{Braaten:1980yq,Drees:1990dq,Gorishnii:1990zu}.
For the Higgs couplings, \textsc{2HDMC} takes the running fermion masses 
through the leading logarithmic corrections to all orders in the $\overline{\rm MS}$ scheme
with the renormalization scale being the mass of the mother Higgs boson.

The first remarkable feature in \fig{fig-BR} is that $\mhf$ almost fixes the branching ratios of $\hf$
although we included all the allowed parameter points.
For $\mhf=80\gev$, $\hf\to\rr$ is the leading decay mode.
For $\mhf=100\gev$, however, $\hf\to W^{(*)} W^*$ is dominant.
If  $\mhf=90\gev$, $\br(\hf\to\rr)\simeq \br(\hf\to W^{(*)} W^*)$.
The branching ratios of $A$ about $\mach$ show a large variation,
except when $\ma$ is near the allowed minimum.
The branching ratios of the charged Higgs boson also show some variation,
but the leading decay mode is almost fixed by $\mach$.
When $\mch$ is below about $\mhf+15\gev$, the dominant decay mode is $\ch\to\tau^\pm\nu$.
We remind the reader that 
$\ch\to\tau^\pm\nu$ associated with two photons
has not been studied in the literature.
If $\mch$ is heavier, $\ch \to \hf W^*$ is the leading decay mode.

\subsection{Constraints from CDF $4\gm X$ and CMS $W\rr$ }
\label{subsec:CDF:CMS}

The public code \textsc{HiggsBounds} covers most results 
of the direct searches for a new particle at the LEP, Tevatron, and LHC.
However, it misses one important search for $\hf$,
the $4\gm X$ mode by the CDF Collaboration~\cite{CDF:2016ybe}.
In addition, the CMS measurement of $pp\to \wpm \rr \to \ell^\pm \nu\rr $~\cite{CMS:2021bvh}
also constrains the model because the final state of $\ell^\pm \nu\rr $
can be generated from $pp \to \hf (\to\rr)\ch(\to \tau^\pm\nu)$
through the leptonic decays of $\tau^\pm$.
In this subsection, we implement the two constraints.

First, we study the consistency with the CDF $4\gm X$ measurement. 
As no evidence of a new signal is observed in the search, 
the CDF Collaboration presented the exclusion plot in the $(\mch,\mhf)$ plane for $\tb=10$ and $\ma=350\gev$.
If we accept Figure 3 of Ref.~\cite{CDF:2016ybe} for $\mch \lsim 150\gev$, 
the entire parameter space for $\mhf=80\gev$
and a large portion of the parameter space for $\mhf=90\gev$ are excluded.
However, the CDF results are based on the assumption that $\br(\hf \to \rr ) \simeq 100\%$ for $\mhf \lsim 95\gev$,
which is not always correct as shown in \fig{fig-BR}.

We find that the effect of $m_{12}^2$ on $\br(\hf\to\rr)$ has not been considered in Ref.~\cite{CDF:2016ybe}.
The partial decay width is~\cite{Djouadi:2005gj}
\bea
\label{eq:Gm:hfrr}
\Gm(\hf \to \rr) = \frac{G_F \,\al_e^2 \, \mhf^3}{128\sqrt{2}\, \pi^3}
\left|
\sba \bar{A}_1 (\tau_W) + \frac{v^2}{2 \mch^2} \lmh_{h H^+ H^-} \, \bar{A}_0 (\tau_{\ch})
\right|^2,
\eea
where $\tau_i = \mhf^2/(4 m_i^2)$. 
The amplitudes $\bar{A}_{1}(\tau)$ and $\bar{A}_{0}(\tau)$ are
\begin{align}
\bar{A}_1 (\tau) &= \tau^{-2}\left[
2\tau^2+3\tau + 3(2\tau-1)f(\tau)
\right],
\\ \nn
\bar{A}_0 (\tau)  &= -\tau^{-2}\left[ \tau - f(\tau) \right],
\end{align}
where $f(\tau) = \arcsin^2 \sqrt{\tau} $ if $\tau<1$.
We define $\bar{A}_{1}(\tau)= - {A}_1^{\mathcal{H}}$
and $\bar{A}_{0}(\tau)= {A}_0^{\mathcal{H}}$ so that $\bar{A}_{1,2}(\tau)>0$,
where the expressions of ${A}_{1,0}^{\mathcal{H}}$ 
are referred to Ref.~\cite{Djouadi:2005gj}.
Since $\sba<0$ in the fermiophobic type-I, 
negative $\lmh_{hH^+ H^-}$ results in the constructive interference between the $\wpm$ and $\ch$ contributions.
So, $ \lmh_{hH^+ H^-} \simeq {\tb (\mhf^2-M^2)}/{v^2} $
implies that large $m_{12}^2$ enhances $\Gm(\hf \to \rr) $.

\begin{figure}[!t]
\centering
\includegraphics[width=0.6\textwidth]{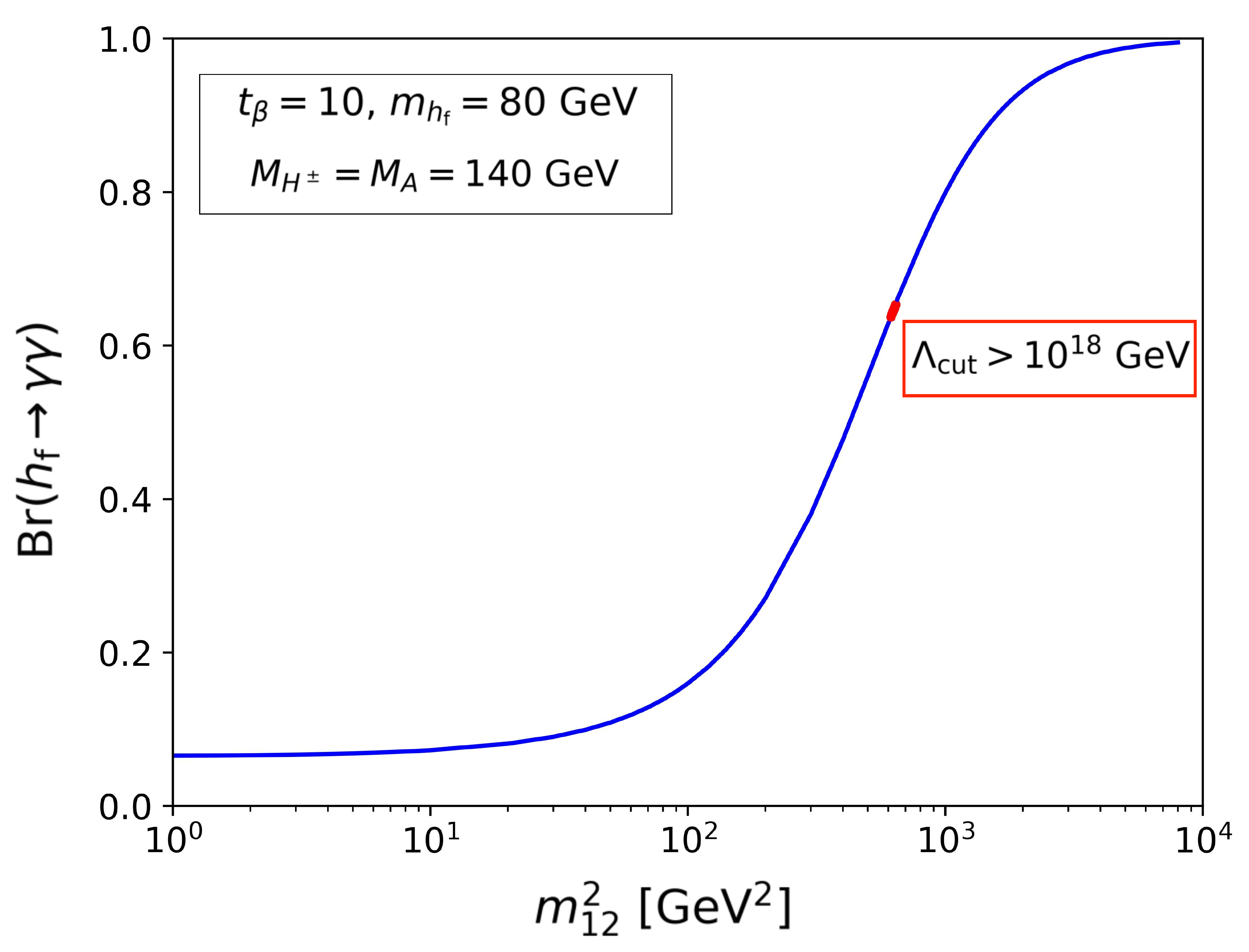}
\vspace{-0.4cm}
\caption{
$\br(\hf\to\rr)$ as a function of $m_{12}^2$ for $\mhf=80\gev$, $\mach=140\gev$, and $\tb=10$.
The red points correspond to the results of the allowed parameter points with $\lmc>10^{18}\gev$.
  }
\label{fig-Brr-m12sq}
\end{figure}

In \fig{fig-Brr-m12sq},
we show the branching ratio of $\hf\to\rr$ as a function of $m_{12}^2$ for 
$\mhf=80\gev$, $\mch=M_A=140\gev$, and $\tb=10$.
It is clearly seen that $\br(\hf\to\rr)$ increases with increasing $m_{12}^2$.
The CDF assumption of $\br(\hf \to \rr ) \simeq 100\%$ for $\mhf \lsim 95\gev$
is valid only when we take a large value of $m_{12}^2$.
Note that the conventional assumption of $M^2 = \ma^2$~\cite{Akeroyd:2003bt}
leads to a large value of $m_{12}^2$,
which yields $m_{12}^2 \simeq 1.9 \times 10^5 \gev^2$ for $\tb=10$ and $\ma=140\gev$.
However, 
the $m_{12}^2$ range allowed by $\lmc>10^{18}\gev$ is considerably small as marked by red points in \fig{fig-Brr-m12sq}.
Therefore, the constraint from the CDF $4 \gm X$ measurement should be reanalyzed 
for the fermiophobic type-I with a high cutoff scale.

\begin{figure}[!t]
\centering
\includegraphics[width=0.6\textwidth]{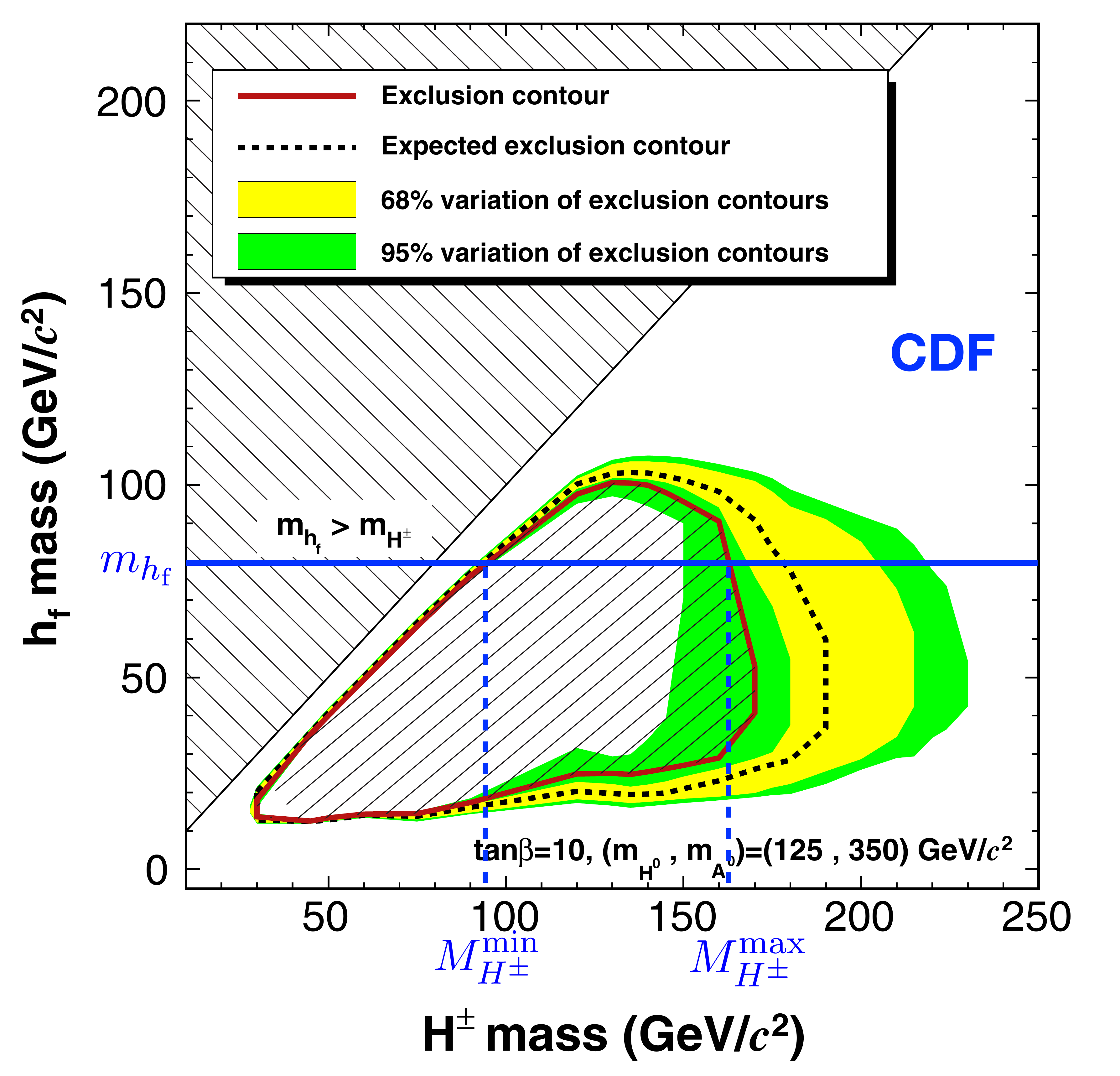}
\vspace{-0.4cm}
\caption{
The CDF exclusion plot of Figure 3 in Ref.~\cite{CDF:2016ybe} for the final selection. 
They set $\tb=10$, and $\ma=350\gev$.
The solid curve is the contour enclosing the exclusion region, the dashed line encloses the median expected exclusion region, and the shaded regions cover the 68\% and 95\% of possible variations of expected contours.
We define $\mch^{\rm min}$ and $\mch^{\rm max}$ for the given $\mhf$ to recast the CDF measurement.
  }
\label{fig-CDF-exclusion}
\end{figure}

Let us describe how we implement the CDF measurement.
Targeting at $p\bar{p}\to \ch(\to \hf \wpm) \hf \to 4\gm +\wpm$,
the CDF Collaboration focused on 
the final state including at least three isolated photons satisfying $E_T^\gm>15\gev$, $|\eta|<1.1$, and $R=0.4$.
Here $E_T$ is the transverse energy.
In the signal region where
two leading photons should satisfy $E_T^{\gm_1} + E_T^{\gm_2} >90\gev$,
ten events were observed.
But none of them contains four photons or a lepton.
Since the observation is consistent with the background expectation,
they provided the exclusion plot over $(\mch,\mhf)$ for $\tb=10$ and $\ma=350\gev$.
For the convenience of the reader,
we present the plot in \fig{fig-CDF-exclusion}.
Since the exclusion applies only to the case of $\br(\hf\to\rr)\approx 100\%$ and $\tb=10$,
we need to recast the CDF result for general $\tb$ and $m_{12}^2$,
which requires the acceptance.
Because the final selection is applied only to two leading photons,
we expect that the dependence of the acceptance on $\mch$ is negligible.
To verify this prediction, we calculated the acceptances for different values of $\mch$
with a given $\mhf$.
The differences in the acceptances are below about 5\%.
For further confirmation,
we calculated $\sg(p\bar{p}\to \hf\ch)  \br(\ch\to \hf W^*)$ 
with a given $\mhf$,
one for $\mch^{\rm min}$ and the other for $\mch^{\rm max}$:
 $\mch^{\rm min}$ and $\mch^{\rm max}$ are defined in \fig{fig-CDF-exclusion}.
We found that two cross sections are almost the same:
for example, the case of $\mhf=80\gev$
yields $\left [\sg\cdot\br(\mch^{\rm min}) - \sg\cdot\br(\mch^{\rm max})\right]/\sg\cdot\br(\mch^{\rm min}) \simeq 3.4\%$.
To efficiently cover the entire parameter points, therefore,
we assume that $\mhf$ alone determines the acceptance of the CDF $4\gm X$ measurement.

 \begin{figure}[!t]
\centering
\includegraphics[width=0.62\textwidth]{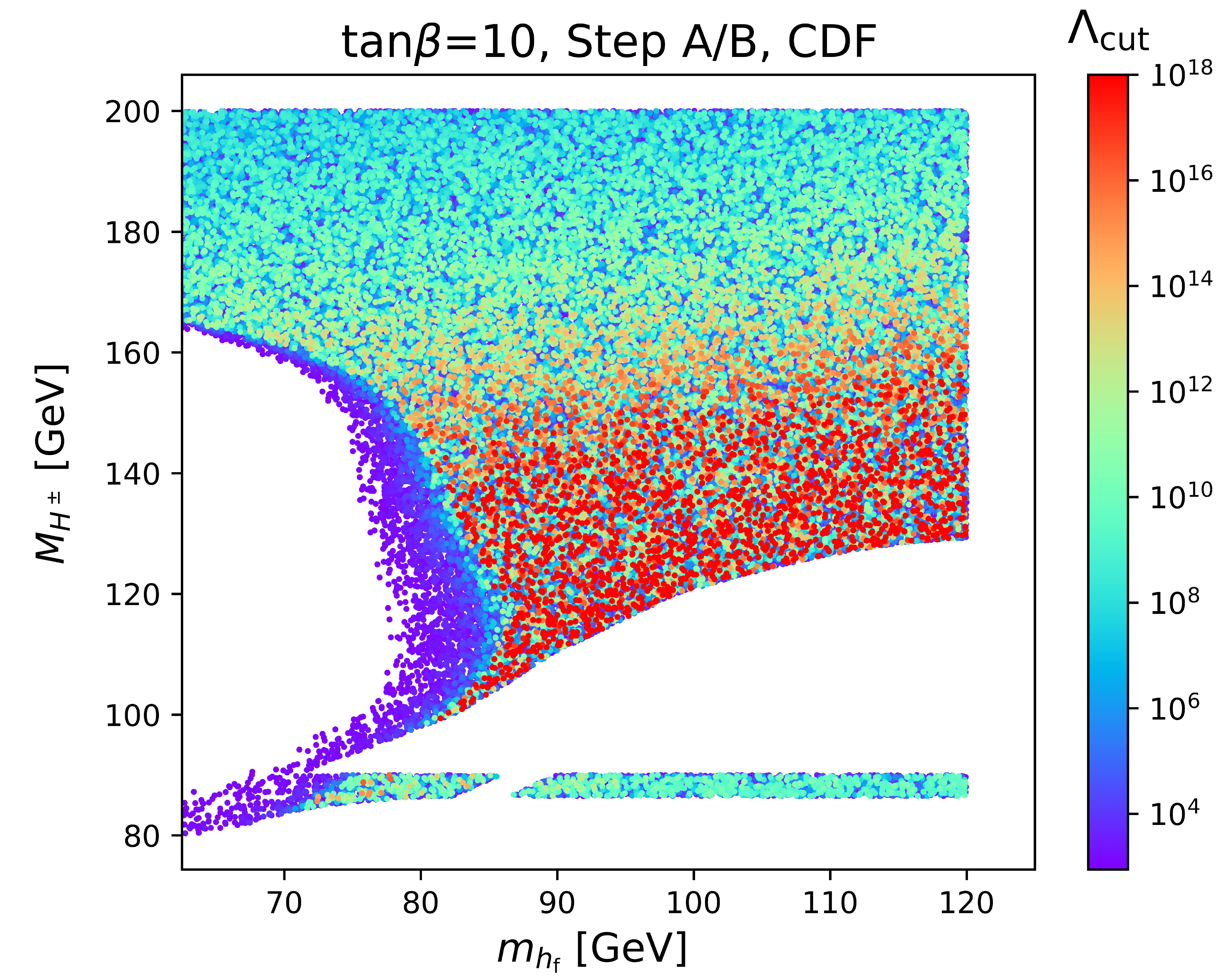}
\vspace{-0.4cm}
\caption{
The parameter points in $(\mhf,\mch)$ allowed by Step A, Step B, $\lmc>10\tev$,
and the CDF $4\gm X$ measurement~\cite{CDF:2016ybe}.
We fix $\tb=10$. 
The color code denotes $\lmc$ in units of GeV.
  }
\label{fig-CDF-McH-mhf-tb10}
\end{figure}

Now
we define the 95\% C.L. upper bound as
\bea
\left(\sg \cdot \br\right)_{\rm CDF}
=  \sg(p\bar{p}\to \hf\ch) \cdot \br(\ch\to \hf W^*) \big|_{\mch=\mch^{\rm min}},
\eea 
where $m_{12}^2$ is chosen to satisfy the CDF assumption of $\br(\hf\to \rr) \approx 100\%$.
For the theoretical prediction in our model,
we calculate $(\sg \cdot \br)_{\rm theory} \equiv \sg(p\bar{p}\to \ch \hf) \br(\ch \to \wpm \hf) \br(\hf\to\rr)^2$
by treating $m_{12}^2$ as a free parameter.
If $(\sg \cdot \br)_{\rm theory}>(\sg \cdot \br)_{\rm CDF}$,
we exclude the parameter point.
In \fig{fig-CDF-McH-mhf-tb10}, we show $\mch$ versus $\mhf$ for $\tb=10$, allowed by the constraints at Step A, Step B, 
and the CDF $4 \gm X$ measurement.
The color code denotes $\lmc$ in units of GeV.
There are important differences between \fig{fig-CDF-exclusion} and \fig{fig-CDF-McH-mhf-tb10}.
In the case of $\mhf=70\gev$, for instance,
the light charged Higgs boson with $\mch \in[80,90] \gev$ is excluded in the CDF analysis
but is permitted with the proper consideration of $\br(\hf\to \rr)$.
However, not a single parameter point for $\mhf=70\gev$ accommodates the Planck cutoff scale.
The CDF $4\gm X$ measurement demands $\mhf \gsim 80\gev$ in the fermiophobic type-I with high cutoff scales.

\begin{figure}[!h]
\centering
\includegraphics[width=1.02\textwidth]{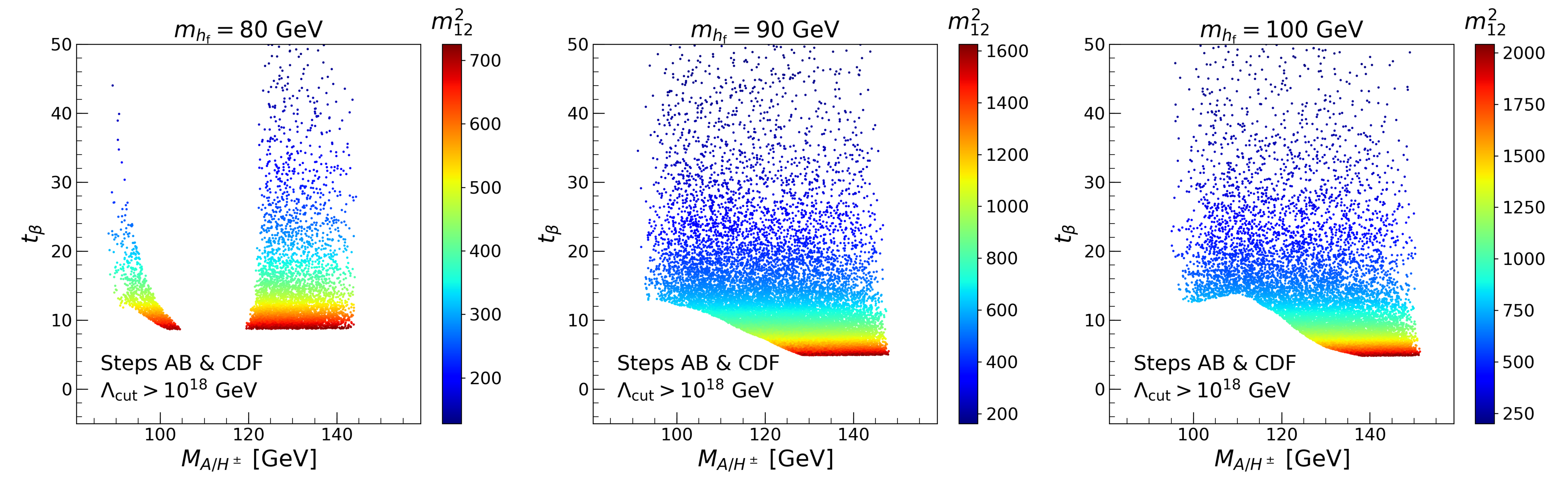}
\vspace{-0.4cm}
\caption{
$\tb$ versus $\mach$ for the parameter points with $\lmc>10^{18}\gev$ after imposing the CDF $4\gm X$ constraint.
The color codes denote $m_{12}^2$ in units of ${\rm GeV}^2$.
The results for $\mhf=80,90,100\gev$ are in the left, middle, and right panel, respectively.
  }
\label{fig-tb-MA-m12sq-Planck-CDF}
\end{figure}

In \fig{fig-tb-MA-m12sq-Planck-CDF},
we show $\tb$ versus $\mach$,
which satisfy Step A, Step B, $\lmc>10^{18}\gev$, and the CDF $4\gm X$ constraint.
The color codes denote $m_{12}^2$ in units of ${\rm GeV}^2$.
The results for $\mhf=80,90,100\gev$ are in the left, middle, and right panels, respectively.
For $\mhf=80\gev$,
the CDF measurement excludes most of the region with $\mch \in [105,119]\gev$,
which is about 60\% of the parameter points that pass Steps A and B.
The range of $\mch \lsim 105\gev$ evades the CDF constraint
because $\br(\ch\to\hf \wpm)$ is small as $\ch\to\tau\nu$ is dominant.
The range of $\mch \gsim 119\gev$ is allowed because the heavy $\mch$ suppresses the production of $p \bar{p}\to \ch\hf$.
The case of $\mhf=90\gev$ ($\mhf = 100\gev$) is weakly (hardly) affected by
the CDF $4\gm X$ measurement.

Another important constraint on $\hf$ is from the diphoton signal.
Basically, the \textsc{HiggsBounds} at Step B checks most of the constraints,
including $\ee\to Z \hf(\to \rr)$~\cite{ALEPH:2002gcw},
$pp\to H \to \hf\hf\to 4\gm$~\cite{Aad:2015bua},
$pp \to H \to \rr bb$~\cite{Aad:2014yja,Khachatryan:2016sey,CMS:2016vpz},
and
$pp \to \hf X \to \rr X$~\cite{ATLAS:2012ad,ATLAS:2012znl,Aad:2014ioa,Khachatryan:2014ira,Khachatryan:2015qba,CMS:2015ocq,Aaboud:2017yyg,Sirunyan:2018aui,ATLAS:2018xad}.
However, the measurement of $pp\to \wpm(\to \ell^\pm \nu) \rr$ at the LHC~\cite{ATLAS:2015ify,CMS:2017tzy,CMS:2021jji} is missing in the \textsc{HiggsBounds}.
The most recent measurement is performed by the CMS Collaboration~\cite{CMS:2021jji}. 
The target events contain an isolated lepton ($\ell^\pm =e^\pm,\mu^\pm)$, 
missing transverse momentum, and two isolated photons.
From the data set with the total integrated luminosity of $137\ifb$,
the cross section for a single lepton flavor under the lepton universality assumption is extracted to be
\bea
\label{eq:CMS:Wrr}
\left. \sg(W_{\ell \nu} \rr )\right|_{\sqrt{s}=13\tev} = 
13.6 \pm 1.9 \,({\rm stat})\pm 4.0 \,({\rm syst}) \pm 0.08 \,({\rm PDF+scale})\fb,
\eea
which is in agreement with the SM prediction at the next-to-leading order.
In the fermiophobic type-I,
the production of $pp\to W^* \to \ch\hf$, followed by $\ch \to \tau^\pm\nu$, $\tau^\pm \to \ell^\pm \nu\nu$, and $\hf \to \rr$,
yields the same final state.
Over the viable parameter points,
we calculated the cross section of $pp \to \ch\hf \to \ell^\pm\nu\nu \rr$ 
with the same selection in Ref.~\cite{CMS:2021jji}
at the detector level by using the \textsc{Delphes} version 3.4.2~\cite{deFavereau:2013fsa}.
The maximum cross section, which occurs for $\mhf=80\gev$,
reaches about $1.6\fb$.
Since it is within the uncertainty in \eq{eq:CMS:Wrr},
the fermiophobic type-I remains unconstrained by the current CMS $W\rr$ measurement.
Nonetheless, we expect that future precision measurement of $W\rr$
will have a significant impact on the model. 

\section{Efficient channels to probe the fermiophobic Higgs boson}
\label{sec:channels}

In the previous section,
we showed that $\lmc>10^{18}\gev$ requires $80\lsim \mhf \lsim 120\gev$ and $90\lsim\mach\lsim 150\gev$.
The mildly compressed mass spectrum for the BSM Higgs bosons makes it challenging to probe the new Higgs bosons at the LHC.
In this section, we pursue efficient discovery channels for the light fermiophobic Higgs boson.

According to whether $\hf$ is produced singly or in pairs,
the target decay modes are different.
For a single production,
$\hf \to \rr$ is certainly the most efficient
because a resonance bump in the invariant mass distribution of two prompt photons is a clean signature of $\hf$.
For the production of two fermiophobic Higgs bosons,
however, $\hf\hf \to \rr W^{*}W^{(*)}$ has more advantages over $\hf\hf \to 4\gm$.
First, $\br(\hf\hf\to \rr W^{*}W^{(*)})$ is larger than $\br(\hf\hf\to 4\gm)$:
\begin{align}
\mhf&=80\gev: & \br(\hf\hf \to \rr WW) &\simeq 36\%, &  \br(\hf \to \rr)^2 &\simeq 35\%,
\\ \nn
\mhf&=90\gev: & \br(\hf\hf \to \rr WW)&\simeq 40\%, &  \br(\hf \to \rr)^2 &\simeq 20\%,
\\ \nn
\mhf&=100\gev: &\br(\hf\hf \to \rr WW) &\simeq 30\%, &  \br(\hf \to \rr)^2 &\simeq 4\%,
\end{align}
where
\bea
\label{eq:brhfhf}
\br(\hf\hf \to \rr WW) = 2 \,\br(\hf\to\rr)\br(\hf\to WW^{(*)}).
\eea
Second, the complication in pairing four photons to trace the prompt $\hf$~\cite{Wang:2021pxc,CMS:2022xxa}
is absent in the $\rr WW$ mode.
Finally, if the production of $\rr WW$ is associated with a $W$ or $Z$ boson,
which happens through $pp\to \ch\hf /A \ch / H^+ H^- $,
the final state can accommodate a same-sign dilepton and two photons,
which enjoys an almost background-free environment.

\begin{table}
  {\renewcommand{\arraystretch}{1.2} 
\begin{tabular}{|c||c|c|}
\hline
\multirow{4}{*}{~~~~~~Target decay modes~~~~~~}& \multicolumn{2}{c|}{$\hf\to2\gm, \quad  \hf\hf \to 2\gm W W^* $ } \\ \cline{2-3}
 &  ~~~~~~Light $\mach$~~~~~~ & ~~~~~~Heavy $\mach$~~~~~~ \\ \cline{2-3}
& $\ch\to \tau^\pm\nu$ & $\ch \to \hf W^*$ \\ \cline{2-3}
 & $A \to bb$ & $A \to \hf Z^*$ \\ \hline\hline
%
{Initial production} & \multicolumn{2}{c|}{Final states} \\ \hline\hline
$q\bar{q}'\to W^* \to \ch\hf $ 
	& $[2\gm]\tau^\pm\nu$ ~\checkmark & $ [ \ellss \rr \met]X$
~\checkmark \\ \hline
$q\bar{q}\to \gm^* /Z^* \to H^+ H^- $  & $ \tau^\pm\nu\tau^\pm\nu$ & $ [ \ellss \rr \met]X$ ~\checkmark
\\ \hline
$q\bar{q}'\to W^* \to A \ch $  & $\bb \tau^\pm\nu$ & $ [ \ellss \rr \met]X$ ~\checkmark
\\ \hline
\end{tabular}
}
\caption{\label{table:production}
The production channels and final states of one or two $\hf$ at the LHC.
The criteria for the light $\mach$ is $\mch \lsim \mhf +15\gev$.
Here $\ellss$ ($\ell^\pm = e^\pm, \mu^\pm $) denotes a same-sign dilepton
and $X$ includes additional leptons and/or jets.}
The particles inside a square
bracket are originated from the decay of $\hf$ or $\hf\hf$. 
The processes with a checkmark are expected to have a high discovery potential at the LHC.
\end{table}

In Table~\ref{table:production},
we summarize the possible production channels of one or two fermiophobic Higgs bosons at the LHC
and the final states from the target decay modes of $\hf\to\rr$ and $\hf\hf\to\rr WW^*$.
We do not consider the processes of $\qq \to Z^* \to A \hf$ and $gg\to AA$ 
because the cross sections are highly suppressed:
$\sg(pp \to A \hf) \lsim 0.03\fb$ and $\sg(pp \to AA) \lsim 1\ab$ at the 14 TeV LHC.
The final state depends on the decays of $\ch$ and $A$, which are determined by $\mach$.
For a light $\mach \lf\lsim \mhf +15\gev \ri$,
the leading decay modes are $\ch\to\tau^\pm\nu$ and $A\to \bb$.
The productions of $\ch\hf$ and $A\hf $ yield $\tau^\pm\nu\rr$ and $\bb\rr$, respectively.
The final state of $\bb\rr$ suffers from huge QCD backgrounds:
the copiously produced jets at the LHC can mimic $b$ quark jets or photons.
The final state of $\tau^\pm\nu \rr$ is expected to have a higher discovery potential:
the $\tau$-tagging and the cut on missing transverse energy help to tame the QCD backgrounds.
For heavy $\mach$, 
the main decay modes are $\ch\to\hf W^*$ and $A\to \hf Z^*$.
The productions of $\ch\hf$, $A\ch$, and $H^+ H^-$ 
can yield the final states of a same-sign dilepton $\ellss$ 
 ($\ell^\pm = e^\pm,\mu^\pm$) and two photons.
In this paper, we focus on 
$\tau^\pm\nu \rr$ and $\ellss \rr \met X$, 
where $X$ includes additional leptons and/or jets.

Let us present the parton-level cross sections of the two final states
over the parameter points satisfying all the constraints and $\lmc>10^{18}\gev$.
Using \textsc{FeynRules}~\cite{Alloul:2013bka},
we first obtained the Universal FeynRules Output (UFO)~\cite{Degrande:2011ua} for the fermiophobic type-I.
After interfacing the UFO file with \textsc{MadGraph5-aMC@NLO}~\cite{Alwall:2011uj},
we computed the cross-sections of $pp \to \ch\hf/A \hf /H^+H^-$ at the 14 TeV LHC.
For the parton distribution function set,
we used \textsc{NNPDF31\_lo\_as\_0118}~\cite{NNPDF:2017mvq}. 
The cross-sections are multiplied by the branching ratios of $\hf$, $A$, and $H^\pm$
obtained from the \textsc{2HDMC}~\cite{Eriksson:2009ws}.

\begin{figure}[!t]
\centering
\includegraphics[width=\textwidth]{fig-Xsec-2rtaunu}\\[-10pt]
\caption{The parton level cross section of $\sg(pp \to \ch\hf\to \tau^\pm\nu\rr)$ at the 14 TeV LHC
about $\mach$
for $\mhf=80\gev$ (left),  $\mhf=90\gev$ (middle), and $\mhf=100\gev$ (right).
The color code denotes $\tb$.
We present the result over the parameter points that satisfy Step A, Step B, the CDF $4\gm X$
constraint, and $\lmc>10^{18}\gev$.}
\label{fig-Xsec-2rtaunu} 
\end{figure}

For the final state of $\tau^\pm\nu \rr$, the parton-level cross section  is
\bea
\sg(\tau^\pm\nu\rr) = \sg(pp \to \ch\hf) \, \br(\ch\to \tau^\pm\nu) \, \br(\hf \to \rr) .
\eea
Figure \ref{fig-Xsec-2rtaunu} shows $\sg(\tau^\pm\nu\rr) $
about $\mach$ with the color code of $\tb$.
The results of  $\mhf=80,90,100\gev$
are respectively in the left, middle,  and right panels.
If $\mach \simeq 100\gev$, the cross sections are sizable,
$\mco(100)\fb$ for $\mh=80,90\gev$ and $\mco(10)\fb$ for $\mhf=100\gev$.
As $\mach$ increases, the off-shell decay of $H^\pm \to W^*\hf$ supersedes $\ch \to \tau^\pm\nu$, 
which reduces $\sg(\tau^\pm\nu\rr)$,
The correlation between the signal rate and $\tb$, 
especially for the heavy $\mach$, is strong such that larger $\tb$, smaller $\sg(\tau^\pm\nu \rr)$.
It is attributed to the Yukawa couplings of $\ch$ being inversely proportional to $\tb$.

The inclusive final state of $\ellss \rr \met X$ comes from three production channels,
$pp \to \ch\hf, A \ch, H^+ H^-$.
Let us first define the following branching ratios for the notational simplicity:
\begin{align}
\label{eq:lep:BR}
\br^W_\ell &= \br(W\to \ell\nu) + \br(W\to \tau\nu)\br(\tau\to \ell\nu\nu),
\\ \nn
\br^Z_{\ell\ell} &= \br(Z \to \elll) , 
\\ \nn \br^Z_{\tau\tau} &= \br(Z\to \ttau), 
\\ \nn
\br^\tau_\ell &=\br(\tau\to \ell\nu\nu).
\end{align}
Then the parton-level cross section from $pp \to \ch\hf$ is
\begin{align}
&\sg(pp \to \ch\hf \to \ellss \rr X) 
\\ \nn
&= 
\sg(pp \to \ch\hf)
\br(\ch\to \hf W^*)\br(\hf\hf \to \rr WW) \lf \br^W_\ell \ri^2,
\end{align}
where $\sg(pp \to \ch\hf) = \sg(pp \to H^+ \hf)+ \sg(pp \to H^- \hf)$
and $\br(\hf\hf \to \rr WW) $ is defined in \eq{eq:brhfhf}.
The second process is from $pp\to H^+ H^- \to \hf W^+ \hf W^-  \to 2\gm W^+ W^-W^+ W^-$.
The sum of two branching ratios of $4W \to \ell^+ \ell^+ X$ and $4W\to \ell^-\ell^- X$ is
\begin{align}
\br^{WWWW}_{\ellss X} &= 2 \lf \br^W_\ell \ri^2 -  \lf \br^W_\ell \ri^4.
\end{align}

The third process is from $pp \to  \ch A \to \rr W^+ W^- \wpm Z$
via  $\ch \to \hf \wpm$, $A  \to  \hf Z $, and $\hf\hf\to 2\gm   W^+ W^- $.
To obtain the branching ratio of the inclusive mode from the long decay chain,
we divide the discussion into two cases: $Z$ decays into non-$\tau$ particles (denoted by $\znotau$) and 
$Z$ decays into $\ttau$ (denoted by $Z_\tau$).
For the case of $\znotau$, let us first discuss the case of $pp \to AH^+ \to \rr W^+ W^- W^+ Z$.
If the $Z$ boson in $  W^+ W^- W^+ Z$ decays into $\elll$,
all the decay modes of $W^+ W^- W^+$
produce $\ell^\pm \ell^\pm$, except for the totally hadronic mode of $WWW\to 6j$.
For the other decays of $Z\to \nnu/q \bar{q}$,
two channels can produce $\ell^\pm \ell^\pm$, $W^+ (\to \ell^+  \nu) W^-(\to \ell^- \nu) W^+(\to \ell^+  \nu)$
and $W^+ (\to \ell^+ \nu) W^-(\to {\rm had}) W^+(\to \ell^+ \nu)$.
Here $W^-(\to {\rm had})$ indicates the non-leptonic decay of $W$, including the hadronically decaying tau lepton.
The branching ratio factor for $\ell^\pm \ell^\pm X$ from $W^+W^- \wpm \znotau$
is
\beq
\br^{WWW Z_{\tiny\mathop{\mbox{non-$\tau$}}}}_{\ell^\pm \ell^\pm X}
=  \br^Z_{\ell\ell} \left[
1-(1-\br^W_\ell )^3
\right] + (1-\br^Z_{\ell\ell}-\br^Z_{\tau\tau})\left[ \lf \br^W_\ell \ri^3 + \lf \br^W_\ell \ri^2 \lf 1-\br^W_\ell \ri \right]
.
\eeq
And the branching ratio factor for $pp\to A H^-$ is the same.
If $Z\to \ttau$,
the decay modes of a pair of tau leptons
(leptonic, semi-leptonic, and hadronic modes)
complicate the calculation of the branching ratio factor for the inclusive same-sign dilepton.
Restricting ourselves to $pp\to H^+ A \to \rr W^+ W^- W^+ Z_{\tau}$,
we find five sources for a lepton, i.e., $W^+ W^- W^+ \tau^+\tau^-$,
which makes $2^5$ cases in total.
The inclusive dilepton final state appears in 20 cases,
which yields the branching ratio of
\bea
\br^{W W W Z_{\tau}}_{\ell^\pm \ell^\pm X}
&=&\br^Z_{\tau\tau}
\Big[
\lf \br^W_\ell \ri^3
+
\lf \br^W_\ell \ri^2 \lf 1-\br^W_\ell \ri
\\ \nn && \qquad
+ 2 \lf \br^W_\ell \ri^2 \lf 1-\br^W_\ell \ri 
\left\{
1-\lf 1-\br^\tau_\ell\ri^2
\right\}
\\ \nn && \qquad
+ 3 \,\br^W_\ell \lf 1-\br^W_\ell \ri^2 
\left\{
1-\lf 1-\br^\tau_\ell\ri^2 - \br^\tau_\ell\lf 1-\br^\tau_\ell\ri
\right\}
\Big].
\eea

In summary, the total parton-level cross section for the $\ellss \rr \met X$ final state
is
\begin{align}
\label{eq:total:Xsec:SS}
&\sg(pp \to \ellss \rr X) 
\\ \nn
&= 
 \sg(pp \to \ch\hf)
\br(\ch\to \hf W^*)\br(\hf\hf \to \rr WW) \lf  \br^W_\ell \ri^2
\\ \nn
&~~+ \sg(pp \to H^+ H^-)
\br(\ch\to \hf \wpm)^2 \,\br(\hf\hf \to \rr WW)\br^{WWWW}_{\ellss X} 
\\ \nn
&~~+ \sg(pp \to A \ch)
\br(A \to \hf Z^*) \br(\ch\to \hf \wpm) \br(\hf\hf \to \rr WW)
\\ \nn
& \qquad \times
\lf \br^{WWW Z_{\tiny\mathop{\mbox{non-$\tau$}}}}_{\ell^\pm \ell^\pm X} +
\br^{W W W Z_{\tau}}_{\ell^\pm \ell^\pm X} \ri ,
\end{align}
where $\sg(pp \to A \ch)=\sg(pp \to A H^+)+\sg(pp \to A H^-)$.

\begin{figure}[!t]
\centering
\includegraphics[width=1.01\textwidth]{fig-Xsec-2rll-separate} 
\caption{$\sg(pp \to \ellss \rr X)$ at the 14 TeV LHC
about $\mach$,
through $pp \to \ch\hf$ (upper panels), $pp \to A\ch$ (middle panels),
and $pp \to H^+ H^-$ (lower panels).
The color code denotes $\tb$.
The results of $\mhf=80,90,100\gev$ are in the left, middle, and right panels, respectively.
The parameter points satisfy Step A, Step B, the CDF $4\gm X$
constraint, and $\lmc>10^{18}\gev$.}
\label{fig-Xsec-2rll-separate} 
\end{figure}

Figure \ref{fig-Xsec-2rll-separate} presents the parton-level cross sections of the $\ellss \rr X$ final state
at the 14 TeV LHC, about $\mach$.
The results through $pp\to \ch\hf$ (upper panels),
$pp \to A \ch$ (middle panels), and $pp\to H^+ H^-$ (lower panels) are separately shown for $\mhf=80,90,100\gev$
in the left, middle, and right panels, respectively.\footnote{Here we present the results only for $\mch>\mhf+15\gev$, which guarantees a sizable $\br(\ch\to \hf W^*)$.
Moreover, 
if the mass of the off-shell $W$ is below a few GeV,
the lepton from $W^* \to \ell\nu$ is too soft to be selected.
}
The largest contribution is from $pp\to \ch\hf$,
of which the maximum cross section is about $10\fb$ for $\mhf=80,90\gev$ and about $5\fb$ for $\mhf=100\gev$.
The minimum cross section is not small, all above about $1\fb$.
The cross section of $pp\to A\ch$ can be also substantial, 
of which the maximum reaches $\mco(1)\fb$.
The pair production of charged Higgs bosons,
which has not been studied for the fermiophobic Higgs boson in the literature, also yields a considerable cross section.
The maximum of $\sg(pp\to H^+ H^- \to \ellss \rr X )$ is of the order of $1\fb$.
Finally, we observe that $\sg(\ellss \rr X)$ in all three production channels increases with $\tb$.
Large $\tb$ suppresses the fermionic decay modes of $A$ and $\ch$,
which enhances the branching ratios of $\ch\to \hf W$ and $A \to \hf Z$.
It is encouraging that the challenging case of $\tb\sim50$ has a higher discovery potential at the LHC.

\begin{figure}[!h]
\centering
\includegraphics[width=1.01\textwidth]{fig-Xsec-2rll-inclusive} \\[-5pt]
\caption{$\sg(pp \to \ellss\rr+X)$ at the 14 TeV LHC
about $\mach$ with the color code of $\tb$.
The results of $\mhf=80\gev$,  $\mhf=90\gev$, and $\mhf=100\gev$,
are in the left, middle, and right panels, respectively.
The parameter points satisfy Step A, Step B, the CDF $4\gm X$
constraint, and $\lmc>10^{18}\gev$.}
\label{fig-Xsec-2rll-inclusive}  
\end{figure}

Figure \ref{fig-Xsec-2rll-inclusive} demonstrates the total cross sections of $\ellss \rr X$ in \eq{eq:total:Xsec:SS},
about $\mach$ with the color code of $\tb$.
The results of $\mhf=80,90,100\gev$ are in the left, middle, and right panels, respectively.
The total cross sections are above $2\fb$ over all the viable parameter points.
Another intriguing point is that the strong correlation between $\sg(\ellss\rr\met X)$ and $\tb$
remains.
The behavior is opposite to that for $\tau^\pm\nu \rr$.
The complementary roles of $\tau^\pm\nu \rr$ and $\ellss \rr X$
is promising to cover the entire parameter space of the model.

\begin{figure}[!h]
\centering
\includegraphics[width=0.47\textwidth]{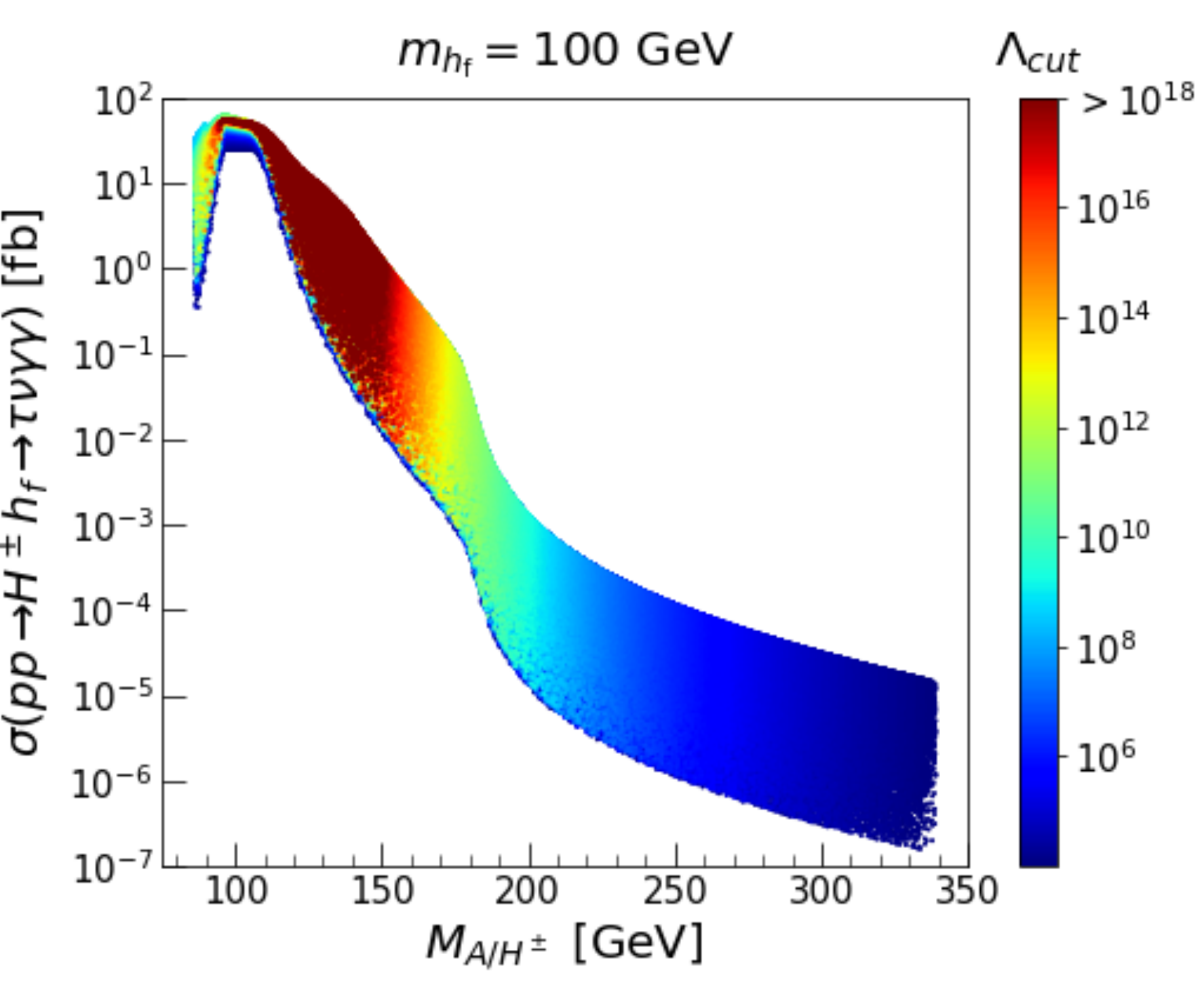}
~~
\includegraphics[width=0.47\textwidth]{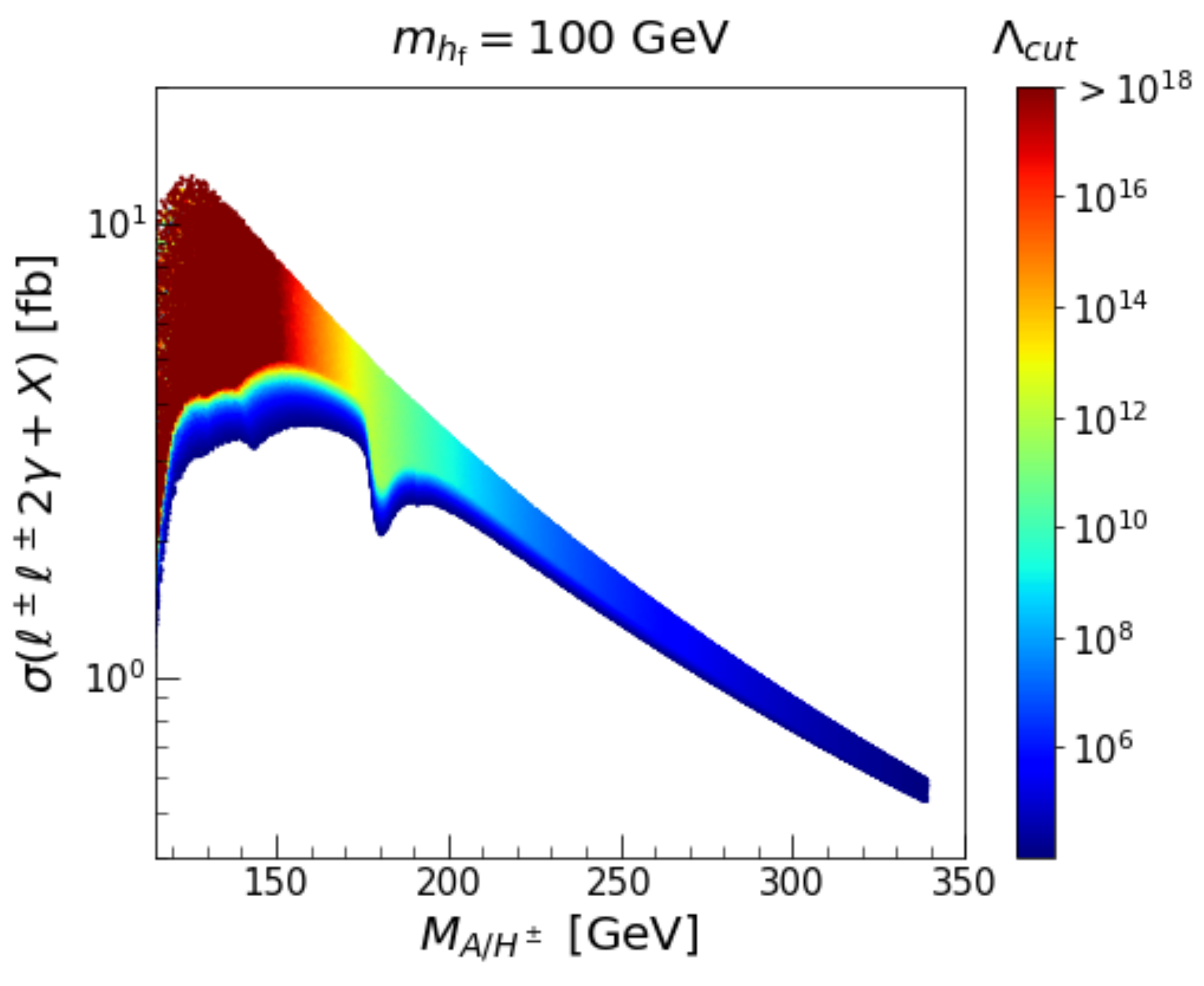} \\[-5pt]
\caption{The total cross sections of
$pp \to \tau^\pm\nu\rr$ (left panel) and $pp \to \ellss \rr+X$ (right panel) at the 14 TeV LHC
about $\mach$ for $\mhf=100\gev$.
The color code denotes the cutoff scale $\lmc$.
The presented parameters satisfy Step A, Step B, the CDF $4\gm X$ measurement,
and $\lmc>10\tev$. }
\label{fig-Xsec-all-cutoff} 
\end{figure}

Finally, let us investigate how the cross sections change 
if we lift the requirement of $\lmc>10^{18}\gev$.
In \fig{fig-Xsec-all-cutoff},
we show the total cross sections of
$pp \to \ch\hf\to \tau^\pm\nu\rr$ (left panel) and $pp \to \ellss\rr X$ (right panel) at the 14 TeV LHC
about $\mach$.
We set $\mhf=100\gev$ and allow $\lmc>10\tev$.
After sorting the parameter points according to $\lmc$,
we stacked them in order of $\lmc$,
putting the points with low $\lmc$ underneath and those with high $\lmc$ on top.
Figure \ref{fig-Xsec-all-cutoff} clearly demonstrates that 
the model with low $\lmc$ has 
a sizable portion of parameter space with the light $\mach \lsim 150\gev$,
which yields similar cross sections to the high $\lmc$ case.
As $\mach$ increases, which is feasible only for a low cutoff scale,
both $\sg(pp \to \tau\nu\rr)$ and $\sg(pp \to \ellss\rr X)$ decrease rapidly.
In summary,
the fermiophobic type-I yields sizable signal rates 
in the final states of $\tau\nu\rr$ and $\ellss\rr X$
if $\mach$ is below about $150\gev$.

\section{Signal-background analysis}
\label{sec:signal:background}
In this section,
we perform the signal-background analysis
for the $\tau^\pm\nu \rr$ and $\ellss\rr \met X$ final states through the detector simulation.
For the Monte Carlo event generation of the signal and backgrounds,
we used the \textsc{MadGraph\_aMC@NLO} version 2.6.7.~\cite{Alwall:2014hca}
with the \textsc{NNPDF31\_lo\_AS.0118} set of parton distribution functions~\cite{Ball:2017nwa}.
For the signal, we adopted the 2HDM \textsc{UFO} file~\cite{Degrande:2011ua}
to generate the productions of $\ch\hf$, $A \ch$, and $H^+ H^-$ using the \textsc{MadGraph}.
The \textsc{Pythia} version 8.243~\cite{Sjostrand:2007gs} was employed 
for the decays of the BSM Higgs bosons in accordance with 
the branching ratio values of 2HDMC~\cite{Eriksson:2009ws}.
The parton showering, hadronization, and hadron decays are also dealt with \textsc{Pythia}.
The fast detector simulation of the signal and backgrounds is carried out 
through the \textsc{Delphes 3.4.2}~\cite{deFavereau:2013fsa}
with the \texttt{delphes\_card\_HLLHC}.
For the jet clustering, we take the anti-$k_T$ algorithm~\cite{Cacciari:2008gp,Mangano:2006rw} 
with the radius parameter $R = 0.4$.
We order the photons and leptons according to the transverse momentum.

A crucial factor in the analysis is the object identification of the photon, tau lepton, and lepton.  
First, the photon identification efficiency is set to the default value in the \textsc{Delphes}, 
which is 95\% if $p_T^\gm>10\gev$.
Since the default setup in the \texttt{delphes\_card\_HLLHC} does not contain the mistagging probability of a QCD jet as a photon,
we additionally impose $P_{j \to \gm} = 5\times 10^{-4}$~\cite{ATLAS:2018fzd}.
Another important issue is the mistagging efficiency of an electron as a photon.
We use the default value,
which ranges from about 1\% to about 10\%, depending on the $p_T^e$.
The tau lepton identification is possible if $\tau^\pm$ decays hadronically.
The hadronic $\tau$-jet differs from QCD jets in that 
it contains a small number of charged and neutral hadrons.
In addition,
when the momentum of $\tau^\pm$ is much larger than $m_\tau$,
the $\tau$-jet is collimated enough to make a localized energy deposit~\cite{CMS:2007sch,Bagliesi:2007qx,CMS:2018jrd}.
Denoting the hadronically decaying $\tau$ by $\tau_{\rm h}$,
we adopt the default values of the tagging and mistagging efficiencies of $\tauh$ in the \textsc{Delphes}, roughly $P_{\tauh\to \tauh} \simeq 0.6$ and $P_{j \to \tauh} \simeq 0.01$.
Note that we are taking a conservative stance for the $\tauh$ identification
since the recent improvements in $\pi^0$ reconstruction and multivariate
discriminants have significantly increased the $\tau_{\rm h}$ tagging efficiency
into $P_{\tauh\to \tauh} \simeq 0.85$~\cite{CMS:2018jrd}.
 
Brief comments on the next-to-leading-order (NLO) corrections are in order here.
In this paper, we take the leading-order results for the signal and backgrounds.
The dedicated calculation at the NLO does not exist 
for the processes of $pp\to\ch\hf/A\ch/H^+H^-$.
It is beyond the scope of this paper for our main purpose to cover the entire parameter space
because the NLO calculation depends on the model parameters.
Nevertheless, a good estimation was performed for $pp\to\ch h_{\rm BSM}$ in Refs.~\cite{Bahl:2021str,Arhrib:2022inj}:
$K\simeq 1.34$ if $\mch$ is light,
where $K$ is the ratio of the NNLO cross section over the LO cross section.
Since the only colored states of the signal processes of $pp\to\ch\hf/A\ch/H^+H^-$ 
are the incoming quarks as in $pp\to\ch h_{\rm BSM}$,
we expect that the $K$-factors of the signal processes are similar to $K\simeq 1.34$.
For the backgrounds, we expect that the $K$-factors are not considerably different 
from the signal $K$-factor because they should include at least one photon or one $W/Z$.
Using \textsc{MadGraph5-aMC@NLO},
we calculated the NLO cross section of the process $pp \to jj\gm$, 
the dominant background for the signal of $\tau^\pm\nu\rr$, 
and found $K_{jj\gm} \simeq 1.25$.
Therefore,
our leading-order results do not overestimate the signal significance.
 
Finally, we use the signal significance including the background uncertainty as~\cite{Cowan:2010js}
\bea
\label{eq:significance}
\mathcal{S} =
\Bigg[2(N_s + N_b) \log \frac{(N_s + N_b)(N_b + \delta_b^2)}{N_b^2 + (N_s + N_b)\delta_b^2} 
 - 
\frac{2 N_b^2}{\delta_b^2} \log\left(1 + \frac{\delta_b^2 N_s}{N_b (N_b + \delta_b^2)}\right)\Bigg]^{1/2},
\eea
where $N_s$ is the number of signal events, $N_b$ is the number of total background events, 
and $\delta_{b} = \Delta_{\rm bg} N_b$ is the uncertainty on the background yields.

\subsection{Final state of $\tau^\pm\nu \rr$} 
\label{subsec:2rtaunu}

Before calculating the signal significance over the entire parameter space,
we need to develop the strategy for the final selection.
So we begin with the following three benchmark points:
\begin{align}
&\hbox{BP-$\tau$1: }& \mhf&=80 \gev, & \mach&=95.8\gev, 
\\ \nn &  & m_{12}^2&=501.1\gev^2, &\tb&=12.5, 
\\[5pt] \nn
&\hbox{BP-$\tau$2: }& \mhf&=90 \gev, & \mach&=100.3\gev, 
\\ \nn & & m_{12}^2&=318.4\gev^2, &\tb&=25.4, 
\\[5pt] \nn
&\hbox{BP-$\tau$3: }& \mhf&=100 \gev, & \mach&=106.9\gev, 
\\ \nn & & m_{12}^2&=274.3\gev^2, &\tb&=36.4. 
\end{align}

The final state of $\tau^\pm\nu \rr$ consists of one $\tauh$-jet, two prompt photons, and missing transverse energy $\met$.
The dominant backgrounds are from $jj\gm$, $j\rr$, $W^\pm \rr$, and $Z\rr$. 
The $jj\gm$ and $j\rr$ backgrounds contribute as the QCD jets are mistagged as a $\tauh$-jet or a photon.
Another important source of photons is photon radiation,
which has been ignored in theoretical studies.
For the basic selection, we impose the followings:
\begin{itemize}
\item[--] We select events with at least one $\tauh$-jet and two leading photons
with $p_{T} > 20\gev$, $|\eta|<2.5$, and the angular separation of $\Delta R =\sqrt{\Delta\eta^2 + \Delta\phi^2} > 0.4$. 
\item[--] We require the missing transverse energy $\met > 20\gev$.
\end{itemize}

Let us describe in more detail how we generated the dominant but tricky background from $jj\gm$.
The first issue about $jj\gm$ is the proportion of the contribution from the photon radiation and 
that from the mistagged jets. 
With the default setting of the \texttt{delphes\_card\_HLLHC} with $P_{j\to \gm}=0$,
only the radiated photons are produced.
If we modify the card with nonzero $P_{j\to \gm}$,
both are produced. 
So,  we generated $6\times 10^6$ events with $P_{j\to \gm}=0$
and another $6\times 10^6$ events with $P_{j\to \gm}=5\times 10^{-4}$.
After the basic selection, the number of events in the former case is 130, 
while that in the latter case is 150. 
Since the photon radiation significantly contributes to the backgrounds,
the simple method adopting $\sg(jj\gm)\times P_{j\to \gm}$ 
underestimates the $jj\gm$ background.

The second difficulty when dealing with $jj\gm$ is the limit on the computation time and cost, 
due to the huge cross section at the event-generation level.
We generated $3\times 10^7$ events of $pp\to jj\gm$
using \textsc{MadGraph5} with the default \texttt{run\_card} requiring $p_T^\gm>10\gev$,
$p_T^j >20\gev$, and $|\eta^{\gm j}|<5$, which is to be called the \texttt{30\,M} events.
Based on the fully showered and hadronized events using \textsc{PYTHIA-8.2},
we carried out the fast detector simulation through the \textsc{Delphes} including $P_{j\to \gm}=5\times 10^{-4}$.
Although 485 events remained after the basic selection, 
not a single event survived when imposing the missing transverse energy cut of $\met>70\gev$.
This is a generic challenge when we encounter an extremely large cross section and an extremely small selection efficiency.

Our solution is to increase selection efficiency 
by using the background events in a more restricted phase space.
Since we will impose stronger cuts in the advanced selection,
the reduced phase space at the parton-level does not affect the background events eventually.
Paying attention that the on-shell decay of $\hf \to \rr$ 
yields the high transverse momentum of the prompt photons,
we generated $2\times 10^7$ events for $jj\gm$, called the \texttt{20\,M} events,
requiring $p_T^{\gm}>60\gev$ at the parton-level. 
Then eight events in the \texttt{20\,M} dataset survived at the final stage.
After confirming that two cross sections from the \texttt{30\,M} and \texttt{20\,M} events match each other
at the first step of the advanced cuts,
we present the cross sections from the \texttt{20\,M} events for the advanced cuts.

\begin{table}[!t]
\begin{center}
{\small
\begin{tabular}{|c||c|c|c||c|c|c|c|c|c|c|}
\toprule
 \multicolumn{8}{|c|}{Cross sections in units of fb for $\tau^\pm\nu \rr$  }\\
\toprule
&BP-$\tau$1 & BP-$\tau$2 & BP-$\tau$3  & \tabincell{c}{$jj\gamma$} & \tabincell{c}{$j\gamma\gamma$}  & \tabincell{c}{ $W^\pm \gamma\gamma$}  &\tabincell{c}{ $Z\gamma\gamma$} \\
\hline  {parton-level with \texttt{MG}} & $197.2$ & $122.1$ & $43.5$  & $7.73\times10^{7}$ & $1.08\times 10^5$ & $140.3$ & $184.7$\\
\hline {Basic Selection} & $21.84$ & $14.87$ & $5.89$  & $1.25\times10^{3}$ & $45.25$ & $0.761$ & $0.954$\\
\hline  \tabincell{c}{$p_{T}^{\gamma_1}>$ 70 GeV \\$p_{T}^{\gamma_2}>$ 40 GeV} & $9.31$ & $7.08$ & $3.11$ & $144.62$ & $28.73$  & $0.205$ & $0.186$\\
\hline  \tabincell{c}{$m_{\gamma_1\gamma_2} \in [62.5,125]\gev$} & $9.20$ & $6.98$ & $3.08$  & $21.94$ & $4.35$ & $0.023$ & $0.032$\\
\hline  \tabincell{c}{$\met>70\gev$ } & $6.49$ & $4.89$ & $2.16$ & $2.51$ & $0.052$ & $0.007$ & $0.003$\\
\hline  \tabincell{c}{veto jets} & $4.36$ & $3.18$ & $1.43$  & $0.98$ & $0.011$& $0.004$ & $0.002$\\
\bottomrule
 \end{tabular}
 }
\caption{The cut-flow of the cross sections (in units of fb) for the final state $\tau^\pm\nu \rr$ and the backgrounds at the 14 TeV LHC.
The parton-level cross sections are obtained 
using the \textsc{MadGraph5} with the default \texttt{run\_card} requiring $p_T^\gm>10\gev$,
$p_T^j >20\gev$, and $|\eta^{\gm, j}|<5$. 
The final step is to veto jets with $p_T>20\gev$ and $|\eta|<2.5$.}
\label{table:Cutflow1}
\end{center}
\end{table}

The advanced cuts consist of the following:
\begin{itemize}
\item[1.] We require that the leading photon has $p_T^{\gm_1} > 70\gev$ and the second leading photon $p_T^{\gm_2}>40\gev$.
\item[2.] The invariant mass of the leading two photons should be inside $[62.5,125]\gev$.
\item[3.] The missing transverse energy should be $\met > 70$ GeV.
\item[4.] We veto an event if it includes the QCD jets, not the $\tauh$-jet, 
with $p_T>20$ GeV and $|\eta|<2.5$. 
\end{itemize}
The condition of $m_{\gamma_1\gamma_2} \in [62.5,125]\gev$ needs an explanation. 
Without the information on $\mhf$ \textit{a priori},
requiring a small mass window for $m_\rr$ is not proper.
For the study of the fermiophobic type-I with high cutoff scales, however,
restricting $m_\rr$ in the permitted $\mhf$ is acceptable.

In Table.~\ref{table:Cutflow1},
we present the cut-flow of the cross sections for the signal and backgrounds.
After the basic selection, 
the cross section of the total backgrounds,
mostly from $jj\gm$, is about two orders of magnitude larger than that of the signal. 
Fortunately, the advanced cuts dramatically suppress the backgrounds:
each cut reduces the $jj\gm$ by an order of magnitude.
At the end of the cut-flow, the dominant backgrounds from $jj\gamma$ are sufficiently tamed. 
For the three benchmark points, we calculate the signal significance with the total integrated luminosity of $300\ifb$ and
 two uncertainties of $\Delta_{\rm bg}=0$ and $\Delta_{\rm bg}=5\%$: 
\begin{align}
\label{eq:significance:taunu}
\Dt_{\rm bg}&=0: & \mathcal{S}_{\rm BP-\tau 1} &=52.8, & \mathcal{S}_{\rm BP-\tau 2} &=41.0,  & \mathcal{S}_{\rm BP-\tau 3} &=20.9,
\\ \nn
\Dt_{\rm bg}&=5\%: & \mathcal{S}_{\rm BP-\tau 1} &=34.2, & \mathcal{S}_{\rm BP-\tau 2} &=27.3,  & \mathcal{S}_{\rm BP-\tau 3} &=14.7.
\end{align}
It is promising that the significances even with an integrated luminosity of $300\ifb$ are well beyond claiming the discovery of $\hf$.

\begin{figure}[!h]
\centering
\includegraphics[width=0.99\textwidth]{fig-detect-Xsec-taunu} 
\caption{The detector-level cross sections of $\sg(pp \to \tau^\pm\nu\rr)$ about $\mach$
after imposing all the advanced cuts, described in the text,
at the 14 TeV LHC. The color code denotes $\tb$.
The results of $\mhf=80,90,100\gev$
are respectively in the left, middle, and right panel.
}
\label{fig-detect-Xsec-taunu} 
\end{figure}

Now let us investigate whether the $\tau^\pm\nu \rr$ final state can probe the entire parameter space.
In \fig{fig-detect-Xsec-taunu}, 
we present the detector-level cross sections of the signal after imposing all the cuts in Table \ref{table:Cutflow1}. 
The results for $\mhf=80,90,100\gev $ are in the left, middle, and right panels, respectively. 
The color codes denote $\tb$. 
In a large portion of the parameter space with light $\mach$,
the cross sections after all the cuts at the detector level are above $1\fb$.
Since the backgrounds are sufficiently tamed,
we do have large enough signal significance.
We present the 3$\sigma$ bound (blue-dashed line) and the 5$\sigma$ bound (red-dashed line) 
for the total integrated luminosity of $300 \ifb$ with the 5\% background uncertainty. 
The straight lines for the 3$\sigma$ and 5$\sigma$ bounds
appear because we applied the same selection criteria,
which is our best bet in the absence of information about $\mhf$ and $\mch$.
The fermiophobic Higgs boson
in the entire parameter space with the light $\mach$ can be discovered through the final state of $\tau^\pm\nu \rr$.

\subsection{Final state of $\ellss\rr\met X$}
\label{subsec:2rssll}

We first focus on the following three benchmark points:
\begin{align}
&\hbox{BP-SS1: }& \mhf&=80 \gev, & \mach&=122.4\gev, 
\\ \nn & & m_{12}^2&=166.5\gev^2, &\tb&=38.4, 
\\[5pt] \nn
&\hbox{BP-SS2: }& \mhf&=90 \gev, & \mach&=112.9\gev, 
\\ \nn &  & m_{12}^2&=166.1\gev^2, &\tb&=48.7, 
\\[5pt] \nn
&\hbox{BP-SS3: }& \mhf&=100 \gev, & \mach&=125.7\gev, 
\\ \nn &  & m_{12}^2&=203.5\gev^2, &\tb&=49.1. 
\end{align}

The inclusive final state consists of two leading leptons with same-sign electric charges (regardless of the flavor),
two isolated photons, and missing transverse energy,
associated with any additional jets and/or leptons.
The backgrounds come from $\wpm Z +$jet, $Z \elll +$jet, and $\wpm Z \gm +$jet.
For the accompanying QCD jet, we include the samples up to one jet
merged with a parton shower using the MLM scheme~\cite{Artoisenet:2012st}.
We do not consider the backgrounds of $t\bar{t}$ and $t\bar{t}V$ 
because they hardly produce the final state of $\ellss \rr \met X$
and can be suppressed by the $b$-jet veto.
Regarding that the leptons and photons in $\ellss \rr \met X$ are soft
because the decays of $H^\pm \to W^*h_f$, $A \to Z^* h_f$, 
and $h_f \to WW^*$ are off-shell,
we take the following basic selection: 
\ben
\item[--] We select events with $N_\gm \geq 2$ and $N_\ell \geq 2$,
where $N_\gm$ ($N_\ell$) is the number of photons (leptons) 
which pass the photon (lepton) criteria in the \texttt{delphes\_card\_HLLHC}~\cite{deFavereau:2013fsa}. 
We further require that two leading leptons have the same-sign charge.
\item[--] Two same-sign leptons and two leading photons must have $p_T>20\gev$. 
In addition, we demand the missing transverse energy to be $\met> 20\gev$.
\item[--]  The two same-sign leptons and the two leading photons should have $|\eta|< 2.5$
and the angular separation between any combination should be $\Dt R >0.4$.
\een

\begin{table*}[!t]
\setlength\tabcolsep{9pt}
\centering
{\renewcommand{\arraystretch}{1.1} 
\small
\begin{tabular}{|c ||c| c| c || c | c|c|}
\toprule
 \multicolumn{7}{|c|}{Cross sections in units of fb for $\ellss \rr \met X$}\\
\toprule
 & BP-SS1 & BP-SS2 & BP-SS3 &  $\wpm Z j$   & $Z\elll j$ & $\wpm Z\gm j$\\
\hline \hline
parton-level with \texttt{MG} & 23.50&  26.95& 12.45 & $1.25 \times 10^3$  & 170.43 & 7.80 \\ \hline
Selecting $\ellss\rr$ & 9.57 & 9.53 & 7.50 & 115.75  & 22.10  & 1.03 \\ \hline
$p_T>20\gev $, 	&	\multirow{2}{*}{1.50} & \multirow{2}{*}{0.77} & \multirow{2}{*}{0.43}  & \multirow{2}{*}{0.164} 	 &\multirow{2}{*}{0.046} & \multirow{2}{*}{0.04} \\
$\met> 20\gev$ & 	&	& 	& &	& 	 \\ \hline
$|\eta|<2.5,~\Dt R>0.4$ & 0.735& 0.354 & 0.227 & 0.070  & 0.027 & 0.021\\
\bottomrule
\end{tabular}
}
\caption{Cut-flow chart of the cross sections (in units of fb) for the final state
  $\rr \ellss \met X$ at the 14 TeV LHC. 
  The parton-level cross sections for the backgrounds include the branching ratios of the leptonic decays of $\wpm$ and $Z$.
}
\label{tab:cutflow:2rssll}
\end{table*}

In Table \ref{tab:cutflow:2rssll},
we present the cut-flow chart of the cross sections of the signal and backgrounds in units of fb.
The results in the first row are the parton-level cross sections by using 
the \textsc{MadGraph5} with the default \texttt{run\_card} and \textsc{NNPDF31\_lo\_as\_0118}~\cite{NNPDF:2017mvq}.
\sblue{
Note that the parton-level cross sections for the backgrounds include the branching ratios of the leptonic decays of $\wpm$ and $Z$:
for example, the cross section for $\wpm Z+$jet is $\sg(pp\to \wpm Z+{\rm jet})\br(\wpm \to \ell^\pm\nu)\br(Z\to \ell^+ \ell^-)$.}
To emphasize the efficiency of each step in the basic selection,
we show the results individually.
First, selecting a same-sign dilepton and a pair of photons considerably suppresses the $\wpm Z +$jet, $\wpm Z \gamma+$jet, and 
$Z\ell^+\ell^- +$jet backgrounds,
eliminating about 91\%, 87\%, and 87\%, respectively.
The selection efficiency of the signal at this level ranges from about 35\% to about 60\%.
Demanding $p_T>20\gev$ for two same-sign leptons and two leading photons as well as $\met>20\gev$
is effective in reducing the backgrounds further.
The survival probability of $\wpm Z +$jet, $\wpm Z \gamma+$jet, and $Z \elll +$jet,
with respect to the previous step,
is about 0.01\%, 0.5\% and 0.03\% respectively.
Finally, the cuts on the rapidity and the angular separation remove almost half of the signal
and the total backgrounds.

For the three benchmark points, we calculate the signal significance with the total integrated luminosity of $300\ifb$.
Considering two uncertainties of $\Dt_{\rm bg}=0$ and $\Dt_{\rm bg}=5\%$,
the significances are
\begin{align}
\label{eq:significance:ss}
\Dt_{\rm bg}&=0: & \mathcal{S}_{\rm BP-SS1} &=23.9, & \mathcal{S}_{\rm BP-SS2} &=13.4,  & \mathcal{S}_{\rm BP-SS3} &=9.3,
\\ \nn
\Dt_{\rm bg}&=5\%: & \mathcal{S}_{\rm BP-SS1} &=21.8, & \mathcal{S}_{\rm BP-SS2} &=12.5,  & \mathcal{S}_{\rm BP-SS3} &=8.7.
\end{align}
%
%
%

\begin{figure}[!h]
\centering
\includegraphics[width=0.99\textwidth]{fig-detect-Xsec-SS} 
\caption{The detector-level cross section of $\sg(pp \to 2\gm\ellss X)$ 
after the basic selection
at the 14 TeV LHC
about $\mach$. The color code denotes $\tb$.
The results of $\mhf=80\gev$,  $\mhf=90\gev$, and $\mhf=100\gev$
are in the left, middle, and right panels, respectively.
}
\label{fig-detect-Xsec-SS} 
\end{figure}

Now let us investigate the discovery potential of the model via $pp \to 2\gm\ellss X$
over the entire parameter space.
In \fig{fig-detect-Xsec-SS},
we present the detector-level cross sections in the cases of $\mhf=80$, 90, $100\gev$
after imposing the basic selection.
The color codes denote $\tb$.
Comparing \fig{fig-Xsec-2rll-inclusive} with \fig{fig-detect-Xsec-SS},
we observe the differences between the parton-level and detector-level cross sections.
The variation of the detector-level cross sections about $\mach$ is smaller than that of the parton-level ones,
which is more prominent for $\mhf=90,100\gev$. 
It is because the lighter $\mach$ results in soft leptons and photons, reducing the selection efficiency.
Consequently, the maximum cross section at the parton level occurs around $\mach \simeq 112\, (123) \gev$
but at the detector level around $\mach \simeq 140 \,(150) \gev$ for $\mhf=90\, (100)\gev$.
For the 5\% background uncertainty, we present the 3$\sigma$ bound (blue-dashed line) 
and the 5$\sigma$ bound (red-dashed line) 
for the total integrated luminosity of $300 \ifb$.
As in the case of $\tau^\pm\nu\rr$,
the same final cuts in Table \ref{tab:cutflow:2rssll} are applied.
It is phenomenal that most of the viable parameter points with heavy $\mach$
can be discovered at $5\sg$ even with the basic selection.

\section{Conclusions}
\label{sec:conclusions}

We have phenomenologically studied the light fermiophobic Higgs boson $\hf$ in the type-I two-Higgs-doublet model,
and proposed two discovery channels of $\tau^\pm\nu \rr$ and $\ellss \rr \met X$
for the first time.
The light $\hf$ is accommodated with the condition of $\al=\pi/2$ in the inverted Higgs scenario 
where the heavier $CP$-even Higgs boson is the observed Higgs boson at a mass of $125\gev$.
Beyond obtaining the parameter points that satisfy the theoretical requirements and the experimental constraints,
we have calculated the cutoff scale $\lmc$ for each parameter point. 
The fermiophobic type-I has a large portion of the parameter space which retains the theoretical stability
(perturbativity, unitarity, and vacuum stability) all the way up to the Planck scale.
The impact of the high cutoff scale is big such that $80\gev \lsim \mhf \lsim 120\gev$ and $90\gev\lsim \mach \lsim 150 \gev$.
In obtaining the viable parameter space,
we have imposed the constraint from the $4\gamma X$ measurement by the CDF Collaboration,
which has played a significant role.
If we include the appropriate branching ratio of $\hf\to \rr$ without any assumption on $m_{12}^2$,
the case with $\mhf=80\gev$ and $\mch\sim100\gev$,
which is excluded in the CDF analysis, is still allowed. 

We have found that the theoretically intriguing parameter region with $\lmc > 10^{18} \gev$
does not meet the conventional assumption of $\br(\hf\to\gamma\gamma)\simeq 100\%$ and $\br(\ch\to \hf \wpm)\simeq 100\%$.
Particularly for the decays of two fermiophobic Higgs bosons,
the decay mode of $\hf\hf\to\gamma\gamma WW$ has a larger branching ratio than $\hf\hf \to4\gamma$.
In addition, the dominant decay mode of the charged Higgs boson is into $\tau^\pm\nu$ for the light $\mch (\lsim \mhf + 15 \gev)$.
Thoroughly investigating the parameter space with high cutoff scales, 
we have proposed two discovery channels that have not been studied in the literature,
 $\tau^\pm\nu \rr$ and $\ellss \rr \met X$.
 The inclusive final state of $\ellss \rr \met X$ consists of a sam-sign dilepton, two prompt photons, and missing transverse energy,
which proceeds through $pp\to \ch\hf$, $pp \to A \ch$, and $pp\to H^+ H^-$.
Furthermore, we showed that even with a lower cutoff, a significant portion of the parameter space can still be probed efficiently using the aforementioned two channels.

The simulation for the signal-background analysis has been performed at the detector level.
The backgrounds from photon radiation as well as the QCD jets mistagged as photons are properly treated.
When calculating the signal significance,
we have taken a conservative stance:
the total integrated luminosity is set to be $300\ifb$, not the expected full luminosity of $3\iab$;
the background uncertainty is included;
the current values of the tagging and mistagging efficiencies for the tau lepton and photon are taken;
the information on $\mhf$, which could restrict the invariant mass of two photons, is not used.
Nevertheless, the results are very encouraging.
The entire parameter space accommodates the significance large enough to claim the discovery.
We urge the experimentalists at the LHC to search for the light fermiophobic Higgs boson
using the final states of $\tau^\pm\nu \rr$ and $\ellss \rr \met X$.

\section*{Acknowledgments}
This work is supported by 
the National Research Foundation of Korea, Grant No.~NRF-2022R1A2C1007583. 

\bibliographystyle{JHEP}
\bibliography{./2high-cutoff-f-phobic}

\end{document}